\documentclass[aoas]{imsart}

\RequirePackage{amsthm,amsmath,amsfonts,amssymb,graphicx,bm}
\RequirePackage[authoryear]{natbib}
\RequirePackage{url}
\RequirePackage{color}
\RequirePackage{comment}
\usepackage[ruled,vlined]{algorithm2e}
\RequirePackage[colorlinks,citecolor=blue,urlcolor=blue]{hyperref}
\startlocaldefs

\usepackage{bbm}
\usepackage{booktabs}
\usepackage{multirow}
\usepackage{threeparttable}
\usepackage[table]{xcolor}

\theoremstyle{plain}

\newtheorem{proposition}{Proposition}

\newtheorem{assumption}{Assumption}
\newtheorem{remark}{Remark}
\newcommand{\expit}{\operatorname{expit}}

\endlocaldefs

\begin{document}

\begin{frontmatter}
\title{Heterogeneous survivor average causal effects beyond monotonicity: Applications to a clinical trial evaluating mechanical ventilation strategies}
\runtitle{Bayesian causal inference for CSACE without monotonicity}

\begin{aug}
\author[A]{\fnms{Zihan}~\snm{Zhu}\ead[label=e1]{zihan.zhu@yale.edu}},
\author[A]{\fnms{Guangyu}~\snm{Tong}\ead[label=e2]{guangyu.tong@yale.edu}},
\author[B]{\fnms{Fernando Godinho}~\snm{Zampieri}\ead[label=e3]{fzampier@ualberta.ca}},
\author[D]{\fnms{Snigdha}~\snm{Jain}\ead[label=e4]{Snigdha.Jain@yale.edu}},
\author[C]{\fnms{Michael O.}~\snm{Harhay}\ead[label=e5]{mharhay@pennmedicine.upenn.edu}},
\author[A]{\fnms{Fan}~\snm{Li}\ead[label=e6]{fan.f.li@yale.edu}},

\address[A]{Department of Biostatistics, Yale School of Public Health%
  \printead[presep={,\ }]{e1,e2,e6}}
\address[B]{Department of Critical Care, University of Alberta%
  \printead[presep={,\ }]{e3}}
\address[C]{Department of Biostatistics, Epidemiology, and Informatics,
  University of Pennsylvania\printead[presep={,\ }]{e5}}
\address[D]{Section of Pulmonary, Critical Care, and Sleep Medicine,
  Department of Internal Medicine, Yale University School of Medicine%
  \printead[presep={,\ }]{e4}}
\end{aug}

\begin{abstract}
Clinical trials in critical care often evaluate outcomes that are truncated by death, such as time to discharge alive, for which treatment effects are not well defined among patients who would not survive under one or both treatment strategies. Principal stratification provides a natural framework for defining survivor causal effects, but existing approaches often rely on monotonicity assumptions that may be implausible in settings where treatment can affect survival in competing directions. This concern is illustrated by the ARDS Network trial of lower versus higher positive end-expiratory pressure (PEEP), in which higher PEEP may benefit some patients by improving alveolar recruitment while harming others through over-distention or hemodynamic compromise. Moreover, the largely null average findings of the original trial do not rule out the possibility of clinically meaningful subgroups that may benefit from or be harmed by higher PEEP. We propose a monotonicity-free framework for identifying conditional survivor average causal effects (CSACE) using an interpretable sensitivity parameter that characterizes latent principal-stratum membership. We then develop a flexible Bayesian estimation strategy based on BART, a posterior-mean-based variable importance measure, and a distribution-aware conditional inference tree that uses the full posterior draws of individualized CSACEs. In simulation studies, the proposed BART-based approach improves individualized CSACE estimation and variable-importance recovery compared with linear modeling, especially under nonlinear treatment-effect heterogeneity. Applied to the ARDS PEEP trial, our method reveals clinically interpretable heterogeneity among estimated always-survivors, identifying subgroups with posterior evidence of benefit or harm that are obscured by the overall null trial results.

\end{abstract}


\end{frontmatter}



\section{Introduction}\label{sec1}

\subsection{The ARDS network randomized trial}
Positive end-expiratory pressure (PEEP) is a central component of mechanical ventilation for patients with acute lung injury and acute respiratory distress syndrome (ARDS). By preventing alveolar collapse at end-expiration, higher PEEP levels may improve arterial oxygenation and reduce ventilator-induced lung injury from repeated opening and closing of non-aerated alveolar units. However, higher PEEP can also be harmful through circulatory depression, increased pulmonary edema, elevated airway pressures, and over-distention-related lung injury~\citep{national2004higher,sahetya2017fifty}. Thus, the relevant clinical question is not simply whether more PEEP is beneficial on average, but for whom the potential benefit of improved recruitment outweighs the potential harm of over-distention and hemodynamic compromise. This benefit--harm tradeoff motivated the ARDS Network trial comparing lower and higher PEEP strategies in mechanically ventilated patients with acute lung injury or ARDS~\citep{national2004higher}. Despite numerous physiology-based approaches proposed to individualize PEEP~\citep{sahetya2017fifty}, no conclusive strategy exists; more than two decades after its publication, the trial's P:F table remains the mainstay of bedside PEEP titration, underscoring the enduring role of randomized trials in shaping clinical practice.

The ARDS trial randomized 549 patients at 23 hospitals to receive either a lower-PEEP or higher-PEEP strategy, with both groups managed under a lung-protective ventilation protocol using a tidal-volume goal of 6 mL per kilogram of predicted body weight and an inspiratory plateau-pressure limit of 30 cm of water. The lower-PEEP arm reflected conventional practice, whereas the higher-PEEP arm tested whether PEEP levels above this conventional range could further improve outcomes within a low-tidal-volume, pressure-limited ventilation strategy. Although the trial achieved clear separation in delivered PEEP and improved oxygenation in the higher-PEEP arm, these physiological gains did not translate into significant clinical improvements on average: mortality before hospital discharge was 24.9\% in the lower-PEEP group and 27.5\% in the higher-PEEP group, and ventilator-free days, ICU-free days, and organ-failure-free days were similar between groups~\citep{national2004higher}.

\begin{figure}[htp!]
\centering
\includegraphics[width=0.85\textwidth]{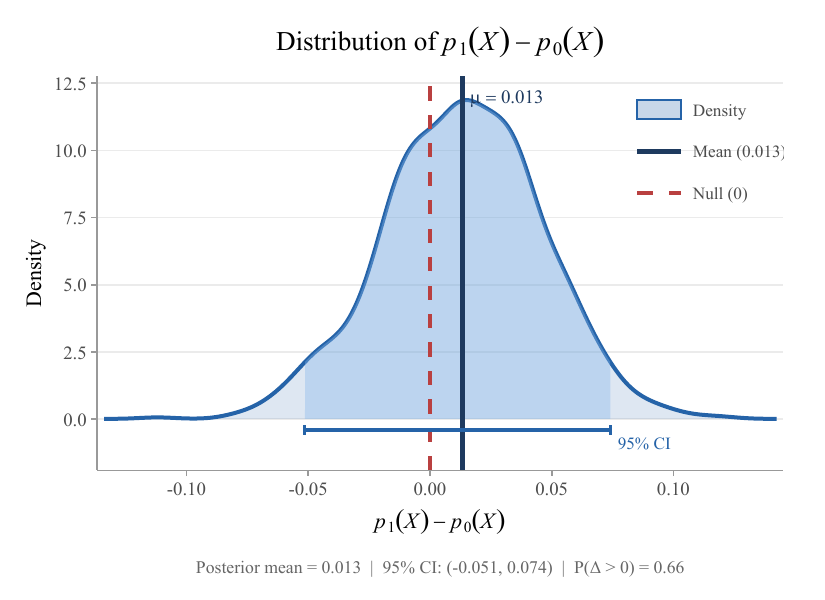}
\caption{
Estimated distribution of the conditional survival contrast under higher versus lower PEEP. The contrast is defined as $p_1(X)-p_0(X)$, where $p_z(X)$ denotes the BART-estimated probability of 60-day survival under treatment arm $z$, with $z=1$ for higher PEEP and $z=0$ for lower PEEP. The dashed red line indicates zero, and the solid dark-blue line indicates the empirical mean. The visible mass below zero indicates that the estimated survival probability under higher PEEP is lower than under lower PEEP for some covariate profiles, providing empirical motivation to avoid a global monotonicity assumption.
}
\label{fig:survival_contrast}
\end{figure}

The clinical assessment of ventilation strategies, however, extends beyond mortality to nonmortality outcomes such as hospital length of stay (h-LOS), which reflect recovery and resource use among patients who survive. Analyzing such outcomes requires confronting truncation by death: for a patient who does not survive, h-LOS is undefined rather than merely missing, so comparing outcomes among observed survivors alone need not target a well-defined causal estimand and can introduce selection bias. Principal stratification offers an effective framework for this problem by classifying patients according to their joint potential survival status under the two strategies, thereby focusing inference on the always-survivor stratum, for whom the nonmortality outcome is well-defined under either treatment strategy~\citep{frangakis2002principal}. Within this framework, existing analyses of causal effects often invoke the monotonicity assumption~\citep{chen2024bayesian}, which in this setting would require that higher PEEP cannot reduce survival for any patient. This restriction is difficult to justify clinically. The trial compared two active ventilation strategies, each with plausible benefits and harms: higher PEEP may improve recruitment and oxygenation, but may also cause over-distention, hemodynamic compromise, or pulmonary edema~\citep{national2004higher,sahetya2017fifty}. Thus, patients who would survive under lower PEEP but not under higher PEEP cannot be ruled out a priori. As an empirical diagnostic, Figure~\ref{fig:survival_contrast} shows the estimated distribution of the conditional survival contrast under higher versus lower PEEP. Although the average contrast is slightly positive, a substantial portion of the distribution lies below zero, suggesting that higher PEEP may be associated with lower estimated survival probability for some covariate profiles. This pattern does not constitute a formal test of monotonicity, but it provides empirical support for treating monotonicity as a questionable assumption in this trial~\citep{tong2025semiparametric}. This consideration motivates our development of a conditional survivor average causal effect framework that does not require the assumption of excluding the harmed principal stratum.

Beyond monotonicity, the trial illustrates another challenge: how to interpret largely null population-level findings when the treatment mechanism suggests competing directions of response. In the original analysis, mortality, ventilator-free days, ICU-free days, and organ-failure-free days were similar between the lower- and higher-PEEP groups, suggesting no clear average clinical benefit of increasing PEEP in a low-tidal-volume, pressure-limited ventilation strategy~\citep{national2004higher}. Previous principal stratification analyses of this trial have similarly found limited evidence of a nonzero average causal effect on survival~\citep{tong2025semiparametric}. However, an average null effect may mask clinically meaningful treatment effect heterogeneity~\citep{munroe2025evidence}. Higher PEEP may benefit patients with more recruitable lungs by improving oxygenation and reducing injury from repeated alveolar opening and collapse, while harming patients who are more vulnerable to over-distention, impaired cardiac output, or pulmonary edema. Thus, the relevant question is not only whether higher PEEP improves outcomes on average, but whether some patients may benefit from higher PEEP while others may be better served by the lower-PEEP strategy. Importantly, this heterogeneity question is inseparable from the truncation problem, since treatment-effect heterogeneity on a non-mortality outcome is only well-defined among patients who would survive under either strategy; this motivates studying heterogeneity within the always-survivor stratum through the conditional survivor average causal effect (CSACE).


\subsection{Related literature and contributions}


By classifying individuals according to their joint potential values of relevant post-treatment variables, such as survival status or treatment receipt, principal stratification focuses attention on subgroups for which the causal comparison is well defined. Within this framework, recent work has developed flexible methods for estimating heterogeneous complier average causal effects in settings with imperfect compliance or instrumental variables~\citep{bargagli2022heterogeneous,johnson2022detecting,spanbauer2024flexible}. More closely related to truncation by death, a broad literature has studied identification, estimation, and sensitivity analysis for survivor average causal effects (SACEs), including approaches based on bounds, Bayesian modeling, semiparametric estimation, and extensions to continuous-time or longitudinal outcomes~\citep{ding2011identifiability,tchetgen2014identification,tong2023bayesian,comment2025survivor,grossi2025bayesian}. In parallel, recent work has begun to move beyond average survivor effects toward heterogeneous or conditional survivor estimands, including CSACE estimation using substitution variables or Bayesian machine learning~\citep{deng2021causal,chen2024bayesian,zhang2026estimation}. These developments address important limitations of analyses that impose monotonicity, which rules out the harmed principal stratum by assuming that treatment cannot reduce survival. However, corresponding identification and sensitivity frameworks for CSACE under nonmonotonicity remain less developed. Moreover, practical considerations for interpreting estimated effect heterogeneity---such as variable importance metrics coupled with principled identification of subgroups---have not been explored in this setting. This gap is important in applications such as the ARDS trial, where the scientific objective is not only to estimate an average effect, but also to understand how that effect varies across baseline covariates.

The most closely related work is the BART-based CSACE analysis of \citet{chen2024bayesian}. Their approach demonstrated the value of Bayesian machine learning for modeling survival and non-mortality outcomes flexibly, allowing nonlinear covariate effects and interactions in both the principal stratum and outcome models~\citep{chipman2010bart}. However, two important gaps remain. First, their analysis relied on the monotonicity assumption, which would have ruled out individuals who would survive under the lower-PEEP strategy but die under the higher-PEEP strategy. This assumption is difficult to justify in the ARDS trial, where higher PEEP may plausibly improve oxygenation for some patients while increasing the risk of over-distention, hemodynamic compromise, or other adverse effects for others. Second, subgroup discovery in existing BART-based CSACE analyses relied on the traditional ``fit-the-fit'' approach, which reduces each individual's posterior CSACE distribution to a point summary, such as the posterior mean, and then fits a regression tree to these point summaries~\citep{hu2021estimating,chen2024bayesian}. Although intuitive and interpretable, this strategy does not fully leverage the posterior information from the Bayesian analysis. A BART-based CSACE analysis yields posterior samples of causal effect for each always-survivor, not merely a single estimated value. Collapsing these posterior distributions before subgroup discovery discards uncertainty, may obscure posterior heterogeneity, and can produce subgroup summaries that are less faithful to the Bayesian inferential target. These limitations motivate an analysis that both relaxes monotonicity and uses the full posterior distribution of individualized CSACEs when selecting covariates, constructing subgroups, and summarizing subgroup-specific benefit or harm.

In light of the above context, our contributions are two-fold. First, we extend the CSACE identification framework beyond the standard monotonicity assumption by introducing an interpretable sensitivity parameter that allows the harmed principal stratum to be present. This enables investigators to assess treatment effect heterogeneity among always-survivors without ruling out by assumption the possibility that treatment may reduce survival for some patients. Second, we develop a BART-based pipeline for CSACE subgroup discovery. We first construct Bayesian estimates of CSACEs to inform variable importance measures (VIM) that identify baseline covariates most relevant to treatment effect heterogeneity among the always-survivors. We then use these covariates to build a posterior-distribution-aware conditional inference tree, leveraging the full posterior samples of CSACEs rather than only posterior summary statistics. The resulting tree serves as a Bayesian summary of heterogeneous survivor causal effects, yielding interpretable subgroups with posterior evidence of benefit or harm while preserving uncertainty from the underlying BART analysis. An implementation is available in the R package \texttt{CSACEtree} at \url{https://github.com/Xu2Zhi1Lan2/CSACEtree}.


\section{Methods}\label{sec2}

\subsection{Principal stratification and the conditional survivor average causal effect}

We consider an independent and identically distributed sample of size $n$, indexed by $i = 1, \dots, n$. Let $Z_i \in \{0,1\}$ denote a binary treatment assignment, where $Z_i = 1$ indicates assignment to the treatment group and $Z_i = 0$ indicates assignment to the control group. Let $D_i \in \{0,1\}$ denote a post-treatment intermediate variable. In the setting of truncation by death, let \(D_i\) denote whether individual \(i\) survives during the follow-up period, with \(D_i=1\) if the individual remains alive so that the final nonmortality outcome is observed and \(D_i=0\) if the individual dies before the final outcome can be measured. Let $Y_i \in \mathbb{R}$ denote the final outcome of interest, and let $\bm X_i \in \mathbb{R}^p$ denote a vector of pre-treatment covariates, potentially high-dimensional. We adopt the potential outcomes framework~\citep{rubin1974estimating,rubin1980randomization}. For each unit $i$, let $Y_i(z)$ and $D_i(z)$, $z \in \{0,1\}$, denote the potential final and intermediate outcomes under treatment assignment $z$, respectively. Under the Stable Unit Treatment Value Assumption (SUTVA), the observed outcomes satisfy
\begin{equation}
Y_i = Z_i Y_i(1) + (1 - Z_i) Y_i(0),
\qquad
D_i = Z_i D_i(1) + (1 - Z_i) D_i(0).
\end{equation}
Principal strata are defined by the joint potential intermediate outcomes~\citep{frangakis2002principal}, as $(D_i(0), D_i(1)) \in \{0,1\}^2$, which induce four subpopulations:
$(1,1)$ (always-survivors),
$(0,0)$ (never-survivors),
$(0,1)$ (the protected), and
$(1,0)$ (the harmed). For a given stratum $(d_0, d_1)$, we define the CSACE as
\begin{equation}
\mu_{11}(\bm x)
=
\mathbb{E}\!\left[
Y_i(1) - Y_i(0)
\;\middle|\;
D_i(0) = 1,\;
D_i(1) = 1,\;
\bm X_i = \bm x
\right].
\end{equation}
The CSACE can be interpreted as the counterpart to the conditional average treatment effect (CATE) in the presence of death truncation. When $D_i$ represents survival status, the final outcome $Y_i(z)$ is well-defined only if the individual survives under assignment $z$, that is, only if $D_i(z) = 1$; for individuals who die before the outcome is measured, $Y_i(z)$ is not merely missing but undefined. Consequently, the individual-level contrast $Y_i(1) - Y_i(0)$ is well-defined only among
always-survivors, who would survive under either assignment, so that this stratum constitutes the largest subpopulation over which a causal contrast in the final outcome is meaningful. Because membership in the always-survivor stratum is determined by the joint potential values $(D_i(0), D_i(1))$ and is thus unaffected by treatment assignment, conditioning on this stratum preserves the causal interpretation of the contrast, as opposed to naive comparisons among observed survivors ($D_i = 1$ in each arm), which condition on a post-treatment event and are generally subject to selection bias. Whereas the survivor average causal effect (SACE), $\mu_{11} = \mathbb{E}[Y_i(1) - Y_i(0) \mid D_i(0) = 1, D_i(1) = 1]$, summarizes the treatment effect for always-survivors as a single scalar \citep{zhang2003estimation}, the CSACE $\mu_{11}(\bm x)$ captures treatment effect heterogeneity within this stratum as a function of pre-treatment covariates \citep{deng2021causal,chen2024bayesian}, enabling a richer characterization of how always-survivor-specific treatment benefits vary across the covariate space.

\subsection{Causal assumptions and point identification}\label{sec:identification}
We introduce the assumptions under which CSACE $\mu_{11}(\bm x)$ is point identifiable.

\begin{assumption}[Treatment Ignorability]
\label{A1}
Treatment assignment is conditionally independent of all potential outcomes given pre-treatment covariates:
\begin{equation}
Z_i \;\perp\!\!\!\perp\;
\{ D_i(0), D_i(1), Y_i(0), Y_i(1) \}
\; \mid \;
\bm X_i.
\end{equation}
\end{assumption}
Assumption~\ref{A1} is guaranteed by randomization in the ARDS trial; covariate adjustment is used to improve efficiency and define conditional effects. In observational studies, it instead requires no unmeasured confounding given \(\bm X_i\)~\citep{imbens2015causal}.

\begin{assumption}[Conditional Principal Ignorability]
\label{A2}
For $z \in \{0,1\}$,
\begin{equation}
\mathbb{E}\!\left[
Y_i(z)
\;\middle|\;
D_i(z)=d_z,\;
D_i(1-z)=d_{1-z},\;
\bm X_i=\bm x
\right]
=
\mathbb{E}\!\left[
Y_i(z)
\;\middle|\;
D_i(z)=d_z,\;
\bm X_i=\bm x
\right].
\end{equation}
\end{assumption}

Principal ignorability assumptions are widely used when principal strata are latent~\citep{jo2009use,ding2017principal,jiang2022multiply}. Assumption~\ref{A2} states that, conditional on $(D_i(z),\bm X_i)$, the mean of $Y_i(z)$ does not further depend on the cross-world survival status $D_i(1-z)$. Unlike treatment ignorability, it is not guaranteed by randomization and requires sufficiently rich baseline prognostic adjustment.

\begin{remark}[Relation between Conditional Principal Ignorability (CPI) and Principal Ignorability (PI)]
\label{remark:CPI_vs_PI}
CPI is weaker than conventional principal ignorability (PI), which posits
\[
\mathbb{E}\!\left[ Y_i(z) \mid D_i(0), D_i(1), \bm X_i \right]
=
\mathbb{E}\!\left[ Y_i(z) \mid \bm X_i \right].
\]
PI removes all mean prognostic information in the principal stratum given $\bm X_i$, whereas CPI permits dependence on the realized $D_i(z)$ and restricts only the additional cross-world dependence on $D_i(1-z)$. Thus, PI implies CPI, but not vice versa.
\end{remark}

\begin{proposition}[Identification of CSACE]
\label{prop:identification}
Under Assumptions~\ref{A1}--\ref{A2} and SUTVA, the conditional survivor average causal effect
\[
\mu_{11}(\bm x)
=
\mathbb{E}\!\left[
Y_i(1) - Y_i(0)
\;\middle|\;
D_i(0) = 1,\;
D_i(1) = 1,\;
\bm X_i = \bm x
\right]
\]
is identified as
\begin{equation}
\label{eq:estimand}
\mu_{11}(\bm x) = m_{11}(\bm x) - m_{01}(\bm x),
\end{equation}
where
\begin{equation}
m_{zd}(\bm x)
=
\mathbb{E}(Y_i \mid Z_i = z,\; D_i = d,\; \bm X_i = \bm x),
\qquad z, d \in \{0,1\}.
\end{equation}
\end{proposition}

Proposition~\ref{prop:identification} does not invoke monotonicity, $D_i(1)\geq D_i(0)$ a.s., and therefore leaves the harmed stratum admissible. The result follows because CPI implies, for $z\in\{0,1\}$,
\[
\mathbb{E}\{Y_i(z) \mid D_i(z)=1,\, \bm X_i=\bm x\}
= \mathbb{E}\{Y_i(z) \mid D_i(0)=1,\, D_i(1)=1,\, \bm X_i=\bm x\},
\]
so the conditional mean among observed survivors in arm $z$ equals the always-survivor mean regardless of the protected and harmed proportions. A proof is provided in Supplementary Section~1.

\begin{assumption}[Principal Score and Conditional Odds Ratio]
\label{A3}
Define the principal score~\citep{ding2017principal}
\[
e_{d_0 d_1}(\bm x)
= \mathbb{P}\!\left\{D_i(0)=d_0,\; D_i(1)=d_1 \;\middle|\; \bm X_i=\bm x\right\},
\qquad d_0, d_1 \in \{0,1\}.
\]
Following recent nonmonotonicity sensitivity frameworks for principal stratification, the dependence structure among principal strata is characterized through the conditional log odds ratio~\citep{tong2025semiparametric}
\begin{equation}
\label{eq:theta}
\theta(\bm x)
= \log\frac{e_{11}(\bm x)\, e_{00}(\bm x)}{e_{01}(\bm x)\, e_{10}(\bm x)},
\end{equation}
which is treated as a sensitivity parameter.
\end{assumption}

The parameter $\theta(\bm x)$ is the conditional log odds ratio relating $D_i(0)$ and $D_i(1)$ and is not identified from observed data. It has a direct regression interpretation: under the logistic model
\begin{equation}
\label{eq:logit_model}
\operatorname{logit}\,\mathbb{P}(D_i(1)=1 \mid D_i(0), \bm X_i=\bm x)
= \alpha(\bm x) + \xi\, D_i(0),
\end{equation}
we have $\theta(\bm x)=\xi$. Sensitivity analysis then varies a single scalar, with larger $\xi$ indicating stronger positive dependence and $\xi=0$ corresponding to conditional independence.

Importantly, while $\theta(\bm x)$ is not identified from observed data, the principal score $e_{11}(\bm x)$ is recoverable once $\theta(\bm x)$ is specified. Let
\begin{equation}
\label{eq::principal_margin}
p_z(\bm x) = \mathbb{P}(D_i=1 \mid Z_i=z,\, \bm X_i=\bm x), \qquad z \in \{0,1\},
\end{equation}
denote the arm-specific survival probabilities, which are directly estimable from observed data. Given $\theta(\bm x)$, the principal score $e_{11}(\bm x)$ is the unique solution to the system
\begin{equation}
\label{eq:principal_score_system}
\begin{cases}
e_{11}(\bm x) + e_{01}(\bm x) = p_1(\bm x), \\
e_{11}(\bm x) + e_{10}(\bm x) = p_0(\bm x), \\
e_{11}(\bm x) + e_{01}(\bm x) + e_{10}(\bm x) + e_{00}(\bm x) = 1, \\
\log\dfrac{e_{11}(\bm x)\, e_{00}(\bm x)}{e_{01}(\bm x)\, e_{10}(\bm x)} = \theta(\bm x).
\end{cases}
\end{equation}
The first three equations constrain the four stratum probabilities to be consistent with observable margins, while the fourth equation encodes the sensitivity parameter. Together they uniquely determine $e_{11}(\bm x)$ as a function of $p_0(\bm x)$, $p_1(\bm x)$, and $\theta(\bm x)$.

For observed survivors, define $\gamma_i=\mathbb{P}\{D_i(0)=1,D_i(1)=1\mid D_i=1,\bm X_i=\bm x\}$. Under Assumption~\ref{A1}, Bayes' theorem gives
\begin{equation}
\label{eq:always_survivor_posterior}
\gamma_i
= \frac{e_{11}(\bm x)}{p_1(\bm x)\,\pi(\bm x) + p_0(\bm x)\,(1-\pi(\bm x))}.
\end{equation}
where $\pi(\bm x)=\mathbb{P}(Z_i=1\mid\bm X_i=\bm x)$. We estimate the marginal always-survivor proportion among observed survivors by averaging $\gamma_i$ over those with $D_i=1$, and classify an observed survivor as likely always-surviving when their posterior probability exceeds this data-driven cutoff. The omitted algebra is given in Supplementary Section~1.

\subsection{Estimation via Bayesian Additive Regression Trees (BART)}
Building on the identification results of Subsection~\ref{sec:identification}, we 
describe the estimation procedure using BART for the three quantities central to our 
framework: the posterior distribution of $\mu_{11}(\bm x)$, the individual-level always-survivor posterior probability, and the estimated always-survivor stratum.

BART is a Bayesian nonparametric ensemble method that represents an unknown
regression function as a sum of many shallow regression trees. For a generic
function \(f(\bm x)\), the BART representation can be written as
\[
f(\bm x)=\sum_{j=1}^{J} h(\bm x;T_j,\mathcal M_j),
\]
where \(T_j\) denotes the structure of the \(j\)th binary tree,
\(\mathcal M_j\) denotes the associated terminal-node parameters, and
\(h(\bm x;T_j,\mathcal M_j)\) returns the terminal-node value assigned to
\(\bm x\) by following the splitting rules in \(T_j\). Priors are placed on
both the tree structures and terminal-node parameters to regularize the fit,
typically favoring shallow trees and small individual-tree contributions.
Thus, each tree acts as a weak learner, while the ensemble can flexibly capture
nonlinearities and interactions among covariates with limited tuning~\citep{chen2024bayesian,chen2025flexible,liu2025bayesian}.

By the identification formula~\eqref{eq:estimand}, the CSACE decomposes as
$\mu_{11}(\bm x) = m_{11}(\bm x) - m_{01}(\bm x)$.
We fit two separate BART models~\citep{chipman2010bart,sparapani2021nonparametric} to estimate $m_{11}(\bm x)$ and $m_{01}(\bm x)$,
using the subsamples $\{i: Z_i=1, D_i=1\}$ and $\{i: Z_i=0, D_i=1\}$,
respectively. For each posterior draw $m$, we compute
\[
\mu_{11}^{(m)}(\bm x) = m_{11}^{(m)}(\bm x) - m_{01}^{(m)}(\bm x),
\]
yielding a collection of $M$ posterior draws
$\{\mu_{11}^{(m)}(\bm X_i): m = 1, \dots, M\}$
for each individual $i$, which characterizes full posterior uncertainty in the
individual-specific CSACE.

To estimate the principal score $e_{11}(\bm x)$, we first obtain the
arm-specific survival probabilities $p_z(\bm x)$
using two Probit BART models~\citep{chipman2010bart,sparapani2021nonparametric} fitted on $\{i: Z_i=z\}$ for $z \in \{0,1\}$,
respectively. This yields posterior draws $\{p_z^{(m)}(\bm x): m=1,\dots,M\}$
for each arm. For each draw $m$ and a specified sensitivity parameter
$\theta(\bm x)$, we substitute $p_0^{(m)}(\bm x)$, $p_1^{(m)}(\bm x)$,
and $\theta(\bm x)$ into system~\eqref{eq:principal_score_system} and solve
for $e_{11}^{(m)}(\bm x)$, propagating posterior uncertainty in the marginal
survival probabilities through to the principal score estimate.

We estimate the propensity score $\pi(\bm x) = \mathbb{P}(Z_i=1 \mid \bm X_i=\bm x)$
via logistic regression. Combining the posterior draws of $e_{11}(\bm x)$
with the estimated $\hat\pi(\bm x)$ and $\hat p_z(\bm x)$, we compute for
each individual $i$ with $D_i=1$ the posterior always-survivor probability
via~\eqref{eq:always_survivor_posterior},
\begin{equation*}
\hat \gamma_i
= \frac{\hat e_{11}(\bm X_i)}
       {\hat p_1(\bm X_i)\,\hat\pi(\bm X_i)
        + \hat p_0(\bm X_i)\,(1-\hat\pi(\bm X_i))}.
\end{equation*}
The population-level cutoff $\mathbb{P}(D(0)=1,\,D(1)=1\mid D=1)$
is estimated by averaging the individual posterior probabilities over all
individuals with $D_i=1$,
\[
\hat{\bar{\gamma}} = \frac{\sum_{i=1}^n D_i
 \hat \gamma_i}{\sum_{i=1}^n D_i}.
\]
Individual $i$ with $D_i=1$ is classified into the always-survivor stratum if
and only if $\hat \gamma_i > \hat{\bar{\gamma}}$. The resulting estimated always-survivor
stratum is $\widehat{\mathcal A}=\{i:D_i = 1,\hat \gamma_i > \hat{\bar{\gamma}} \}$.

\subsection{Variable importance measure and subgroup discovery via distribution-aware conditional inference tree}

Having obtained the CSACE posterior draws for all individuals and the estimated
always-survivor stratum $\widehat{\mathcal A}$, we now describe a principled two-stage strategy for subgroup discovery. The first stage introduces a
variable importance measure (VIM)~\citep{hines2025variable,ziersen2025variable} for CSACE to identify covariates that
are genuinely informative for treatment effect heterogeneity. The second stage uses these selected covariates to grow
a distribution-aware conditional inference tree on $\widehat{\mathcal A}$.
The VIM uses the posterior mean CSACE surface to screen covariates within
$\widehat{\mathcal A}$, consistent with the survivor-specific target defined
below. DaCIT subsequently uses the full posterior draws within the same
estimated stratum to construct an interpretable partition. Thus, the VIM and
DaCIT provide complementary, exploratory summaries of the estimated CSACE
surface, but only the tree is distribution-aware.

Define the survivor-specific prediction risk
\[
\mathcal{L}(f)
= \mathbb{E}\!\left[
  \bigl(\mu_{11}(\bm X_i) - f(\bm X_i)\bigr)^2
  \;\middle|\; D_i(0)=1,D_i(1) = 1
  \right].
\]
For a covariate subset $S \subset \{1,\dots,p\}$, define the reduced CSACE
\[
\mu_{11,S}(\bm x)
= \mathbb{E}\!\left[
  \mu_{11}(\bm X_i)
  \;\middle|\; \bm X_{i,-S} = \bm x_{-S},\; D_i(0)=1,D_i(1) = 1
  \right].
\]
The CSACE variable importance measure (CSACE-VIM) for subset $S$ is
\[
\Theta_S
= \mathbb{E}\!\left[
  \bigl(\mu_{11}(\bm X_i) - \mu_{11,S}(\bm X_i)\bigr)^2
  \;\middle|\; D_i(0)=1,D_i(1) = 1
  \right],
\]
which admits the equivalent variance decomposition
\[
\Theta_S
= \mathrm{Var}\!\left(\mu_{11}(\bm X_i) \mid D_i(0)=1,D_i(1) = 1\right)
- \mathrm{Var}\!\left(\mu_{11,S}(\bm X_i) \mid D_i(0)=1,D_i(1) = 1\right).
\]

\begin{remark}[LOO versus KOI importance for CSACE]
\label{remark:vim}
Within the always-survivor population $D_i(0)=1,D_i(1) = 1$, two natural notions of variable
importance arise. The leave-one-out (LOO) importance for covariate $X_j$,
$\Theta_{\{j\}}$, quantifies the reduction in unexplained heterogeneity of
$\mu_{11}(\bm X_i)$ when $X_j$ is excluded while all remaining covariates
are retained, capturing the conditional contribution of $X_j$ to CSACE
heterogeneity. The keep-one-in (KOI) importance measures the heterogeneity
explained by $X_j$ alone, and while useful for initial screening, may be
inflated under strong covariate correlations. LOO and KOI therefore
characterize conditional necessity and marginal sufficiency of covariates
in explaining CSACE heterogeneity, respectively.
\end{remark}

\begin{algorithm}[ht]
\caption{Estimation of CSACE-VIM}
\label{alg:csace-vim}
\begin{enumerate}
\item Fit the outcome BART models using the corresponding observed-survivor
samples and obtain the posterior mean estimates
\[
\hat m_{11}(\bm x) = \mathbb{E}[m_{11}(\bm x) \mid \text{data}],
\qquad
\hat m_{01}(\bm x) = \mathbb{E}[m_{01}(\bm x) \mid \text{data}],
\]
For each $i\in\widehat{\mathcal A}$, construct the plug-in CSACE estimate
\[
\hat\mu_{11}(\bm X_i) = \hat m_{11}(\bm X_i) - \hat m_{01}(\bm X_i),
\qquad i\in\widehat{\mathcal A}.
\]

\item Among individuals in $\widehat{\mathcal A}$, regress
$\hat\mu_{11}(\bm X)$ on the full covariate vector $\bm X$ using BART to
obtain the full projection
$\hat\mu_{11}^{\mathrm{full}}(\bm X)$,
and regress $\hat\mu_{11}(\bm X)$ on the reduced covariate vector
$\bm X_{-S}$ to obtain the reduced projection $\hat\mu_{11,S}(\bm X)$.

\item Estimate the CSACE-VIM by
\[
\hat\Theta_S
= \frac{1}{|\widehat{\mathcal A}|}
  \sum_{i\in\widehat{\mathcal A}}
  \Bigl[
  \bigl(\hat\mu_{11}(\bm X_i) - \hat\mu_{11,S}(\bm X_i)\bigr)^2
  -
  \bigl(\hat\mu_{11}(\bm X_i) - \hat\mu_{11}^{\mathrm{full}}(\bm X_i)\bigr)^2
  \Bigr].
\]
\end{enumerate}
\end{algorithm}
Based on $\hat\Theta_S$, covariates are ranked by their estimated VIM, and
the $K$ highest-ranked covariates are retained to form the active covariate
set $\widehat{\mathcal{S}}$ for subsequent subgroup discovery, where $K$ is
specified before tree construction.

The second stage of our subgroup discovery procedure grows a
distribution-aware conditional inference tree (DaCIT) on the estimated
always-survivor stratum $\widehat{\mathcal{A}}$, using only the covariates
in $\widehat{\mathcal{S}}$ selected by the VIM screening step. DaCIT extends the conventional scalar-response tree framework by treating the individual-specific posterior distributions of $\mu_{11}(\bm x)$ as distribution-valued responses. Rather than reducing the posterior draws to a point summary such as the posterior mean, DaCIT uses the full posterior distribution to guide both variable selection and split determination at each node. Consequently, subgroup discovery can reflect differences not only in the estimated level of CSACE, but also in posterior dispersion, skewness, and other distributional features that may be obscured by mean-based comparisons alone.

DaCIT retains the central principle of conditional inference trees, which separate variable selection from split-point determination within a recursive hypothesis-testing framework \citep{hothorn2006unbiased}. At each node, candidate covariates are first tested for association with the response; the most strongly associated covariate is selected, and its optimal split is then determined using a separate objective criterion. This separation reduces the variable-selection bias that can arise in conventional CART procedures.

To assess the association between covariates and a distribution-valued response, we represent the individual-specific posterior distributions through their pairwise probability distances. For each individual $i \in \widehat{\mathcal{A}}$, let \[ \mathcal{M}_i = \left\{ \mu_{11}^{(m)}(\bm X_i):m=1,\ldots,M \right\} \] denote the posterior sample of the individualized CSACE, and let $\widehat f_i$ be the density estimated from $\mathcal{M}_i$. We define $\delta_{ij} = d_{\mathcal P}\!\left(\widehat f_i,\widehat f_j\right)$, $i,j\in\widehat{\mathcal A}$, where $d_{\mathcal P}$ is a probability metric, and collect these distances in \[ \bm\Delta = (\delta_{ij})_{i,j\in\widehat{\mathcal A}}. \] Possible choices of $d_{\mathcal P}$ include the total variation, Hellinger, and Jensen--Shannon distances \citep{gibbs2002choosing,lin1991divergence}. Unlike a distance based only on posterior means, these metrics compare the full posterior distributions and can therefore reflect differences in location, dispersion, skewness, and other distributional features.

At a node $t$ with index set $\mathcal I_t$, DaCIT extracts the distance submatrix $\bm\Delta_t$ and its Gower-centered representation
\[
\bm G_t=-\tfrac12\bm J_t\bm\Delta_t^{\circ2}\bm J_t,
\qquad
\bm J_t=\bm I_{n_t}-n_t^{-1}\bm1_{n_t}\bm1_{n_t}^{\top},
\]
where $n_t=|\mathcal I_t|$~\citep{anderson2001new}. For each available covariate, a marginal PERMANOVA test evaluates its association with $\bm\Delta_t$ while adjusting for the other candidates. The node is terminal if no $p$-value is below $\alpha$; otherwise DaCIT selects $j_t^*=\arg\min_j p_{t,j}$.

More specifically, let $\bm H_t^{\mathrm{full}}$ and $\bm H_{t,-j}$ be the projection matrices for the full candidate set and the reduced set excluding $X_j$. Define
\[
SS_{t,j}=\operatorname{tr}\{(\bm H_t^{\mathrm{full}}-\bm H_{t,-j})\bm G_t\},
\qquad
SS_{t,E}=\operatorname{tr}\{(\bm I_{n_t}-\bm H_t^{\mathrm{full}})\bm G_t\}.
\]
The marginal statistic
\[
F_{t,j}=\frac{SS_{t,j}/q_{t,j}}{SS_{t,E}/(n_t-r_t)}
\]
uses $q_{t,j}=\operatorname{rank}(\bm H_t^{\mathrm{full}})-\operatorname{rank}(\bm H_{t,-j})$ and $r_t=\operatorname{rank}(\bm H_t^{\mathrm{full}})$; its $p$-value is obtained by permutation. This adjustment separates covariate selection from the subsequent search over cutpoints.

For each admissible split $c$ of $X_{j_t^*}$, let $z_{t,c,i}=\mathbbm1(X_{ij_t^*}\in\mathcal L_{t,c})$ indicate membership in the left child, where $\mathcal L_{t,c}=\{x:x\leq c\}$ for a continuous covariate and is a candidate subset of levels for a categorical covariate. Centering $\bm z_{t,c}$ gives $\widetilde{\bm z}_{t,c}=\bm J_t\bm z_{t,c}$ and the rank-one projection
\[
\widetilde{\bm H}_{t,c}=\widetilde{\bm z}_{t,c}
(\widetilde{\bm z}_{t,c}^{\top}\widetilde{\bm z}_{t,c})^{-1}
\widetilde{\bm z}_{t,c}^{\top}.
\]
DaCIT selects
\[
c_t^*=\arg\max_c Q_t(c),\qquad
Q_t(c)=
\frac{\operatorname{tr}(\widetilde{\bm H}_{t,c}\bm G_t\widetilde{\bm H}_{t,c})}
{\operatorname{tr}\{(\bm I_{n_t}-\widetilde{\bm H}_{t,c})\bm G_t(\bm I_{n_t}-\widetilde{\bm H}_{t,c})\}},
\]
which maximizes between-child separation relative to within-child distributional variation. Because candidate splits at a node have the same degrees of freedom, maximizing $Q_t(c)$ is equivalent to maximizing the associated PERMANOVA pseudo-$F$ statistic. The selected covariate is removed from both child-node candidate sets, yielding a parsimonious hierarchy. Further implementation details and equivalent trace representations are given in Supplementary Section~3.

\begin{algorithm}[htpb!]
\caption{Distribution-Aware Conditional Inference Tree (DaCIT) for CSACE Subgroup Discovery}
\label{alg:csace-subgroup}
\begin{enumerate}
\item \textbf{Input.} Posterior draws $\{\mu_{11}^{(m)}(\bm X_i): m=1,\dots,M\}$
for each $i \in \widehat{\mathcal{A}}$, and active covariate set
$\widehat{\mathcal{S}}$ from Algorithm~\ref{alg:csace-vim}.

\item \textbf{Distributional representation.} Estimate each posterior density $\hat f_i$ and construct $\bm\Delta=(d_{\mathcal P}(\hat f_i,\hat f_j))$. Initialize the root with all $i\in\widehat{\mathcal A}$ and $\mathcal V^{(0)}=\widehat{\mathcal S}$.

\item \textbf{Recursive splitting.} At node $t$ with distance submatrix $\bm\Delta_t$:
\begin{enumerate}
\item terminate if the maximum depth is reached; otherwise compute the Gower-centered matrix $\bm G_t$ and the marginal PERMANOVA $p$-values;
\item terminate if all $p$-values exceed $\alpha$; otherwise select $j_t^*=\arg\min_j p_{t,j}$;
\item select the admissible split maximizing $Q_t(c)$, create two child nodes, and set $\mathcal V^{(t+1)}=\mathcal V^{(t)}\setminus\{X_{j_t^*}\}$.
\end{enumerate}

\item \textbf{Output.} Terminal-node subgroups within $\widehat{\mathcal A}$ with distinct posterior CSACE distributions.
\end{enumerate}
\end{algorithm}

\begin{remark}[Interpretation and inferential scope]
\label{remark:no-double-dipping}
The VIM and DaCIT are constructed from CSACE estimates obtained using the same
data. The VIM summarizes covariate relevance using the posterior mean CSACE,
whereas DaCIT summarizes differences among the full individualized posterior
distributions. The permutation-based $p$-values used to guide tree splitting
should therefore not be interpreted as confirmatory tests that are independent
of the initial BART fit or the VIM screening step. The selected variables,
tree structure, and terminal-node profiles are exploratory summaries that
require external validation before clinical use.
\end{remark}

\section{Simulation}

We considered data-generating processes designed to evaluate recovery of heterogeneous survivor causal effects and their covariate drivers under nonlinear principal-stratum and outcome models. For each sample size $n$, we generated independent observations $\{(\bm X_i,Z_i,D_i,Y_i)\}_{i=1}^n$, where $\bm X_i=(X_{i1},\ldots,X_{ip})^\top$ and $X_{ij}\stackrel{\mathrm{i.i.d.}}{\sim}\mathrm{TN}(-10,10;0,1)$. Only the first five covariates entered the principal-stratum and outcome models; the remaining covariates were noise variables. Consistent with the motivating randomized trial, treatment was generated as $Z_i\sim\mathrm{Bernoulli}(0.5)$ independently of $\bm X_i$.

Let $(D_i(0),D_i(1))$ denote the potential survival indicators. Their marginal probabilities, $p_z(\bm x)=\Pr\{D_i(z)=1\mid\bm X_i=\bm x\}$, were nonlinear functions of the first five covariates. The four principal-stratum probabilities were then determined by these margins and the conditional odds ratio $\omega=\exp(\xi)$, with $\omega=1$ denoting conditional independence and $\omega=\infty$ denoting the monotonicity-limit construction. This design allowed nonlinear and nonadditive principal-stratum assignment while retaining a transparent dependence parameter. Complete functional forms and parameter values are provided in Supplementary Section~2.

Potential outcomes were generated from nonlinear surfaces involving the same five covariates, with stratum dependence constructed to satisfy conditional principal ignorability. The resulting true CSACE, $\mu_{11}(\bm x)=m_{11}(\bm x)-m_{01}(\bm x)$, was known and nonlinear, providing the ground truth for evaluating individualized CSACE estimation and VIM recovery. Complete outcome-generating formulas are provided in Supplementary Section~2.

\subsection{Experiment 1: Estimation of the CSACE Surface}
\label{sec:sim-csace-estimation}

The first experiment evaluated estimation of the conditional survivor average
causal effect (CSACE) surface, $\mu_{11}(\bm X_i)$. We considered sample sizes
$n\in\{100,500\}$, covariate dimensions $p\in\{5,20\}$, and principal-stratum
odds ratios $\omega\in\{1,1.5,\infty\}$, where larger values of
$\omega$ correspond to stronger positive dependence between $D(0)$ and $D(1)$
and $\omega=\infty$ represents the monotonicity-limit construction. For each
simulated data set, we estimated $\mu_{11}(\bm X_i)$ using the proposed
identification formula, with the nuisance outcome regressions estimated either
by BART or by a parametric linear regression model. The linear model serves as
a benchmark for assessing the value of flexible nonlinear outcome modeling.

Let $e_{ri}=\widehat\mu_{11,r}(\bm X_{ri})-\mu_{11}(\bm X_{ri})$ denote the
individual-level estimation error for individual $i$ in Monte Carlo replicate
$r$, and let $R$ be the number of replicates. We summarized performance across
replicates by
\[
\operatorname{Bias}
=
R^{-1}\sum_{r=1}^R n^{-1}\sum_{i=1}^n e_{ri},
\qquad
\operatorname{RMSE}
=
\left\{
R^{-1}\sum_{r=1}^R n^{-1}\sum_{i=1}^n e_{ri}^2
\right\}^{1/2}.
\]
For each replicate, Precision in Estimation of Heterogeneous Effect
(PEHE) is computed as the root mean squared individual-level effect error. We report its
Monte Carlo average,
\[
\operatorname{PEHE}
=
R^{-1}\sum_{r=1}^R
\left\{n^{-1}\sum_{i=1}^n e_{ri}^2\right\}^{1/2},
\]
which directly measures recovery of the heterogeneous CSACE surface~\citep{hill2011bayesian,shalit2017estimating}. The distinction is the order of aggregation: RMSE pools squared errors across replicates before taking the square root, whereas PEHE averages the replicate-specific root errors. To assess directional decision error, we also report
\[
\operatorname{Regret}
=
n^{-1}\sum_{i=1}^n
|\mu_{11}(\bm X_i)|
\mathbb I\{
\operatorname{sign}(\widehat\mu_{11}(\bm X_i))
\neq
\operatorname{sign}(\mu_{11}(\bm X_i)),
\ \mu_{11}(\bm X_i)\neq 0
\},
\]
which measures the magnitude-weighted fraction of individuals for whom the
estimated direction of the conditional treatment effect is incorrect.

\begin{table}[htpb!]
\centering
\begingroup
\fontsize{10}{12}\selectfont
\begin{threeparttable}
\caption{Performance of BART and linear regression for estimating the CSACE surface in Experiment 1.}
\label{tab:sim1}
\begin{tabular}[t]{rrcrrrrrrrr}
\toprule
\multicolumn{3}{c}{ } & \multicolumn{4}{c}{BART} & \multicolumn{4}{c}{Parametric linear regression} \\
\cmidrule(l{3pt}r{3pt}){4-7} \cmidrule(l{3pt}r{3pt}){8-11}
$p$ & $n$ & $\omega$ & Bias & RMSE & PEHE & Regret & Bias & RMSE & PEHE & Regret\\
\midrule
\addlinespace[0.4em]
\multicolumn{11}{l}{$p=5,\ n=100$}\\
\cellcolor{gray!8}{\hspace{1em}5} & \cellcolor{gray!8}{100} & \cellcolor{gray!8}{1} & \cellcolor{gray!8}{0.334} & \cellcolor{gray!8}{1.982} & \cellcolor{gray!8}{1.957} & \cellcolor{gray!8}{0.356} & \cellcolor{gray!8}{0.351} & \cellcolor{gray!8}{2.352} & \cellcolor{gray!8}{2.321} & \cellcolor{gray!8}{0.609}\\
\hspace{1em}5 & 100 & 1.5 & 0.331 & 1.977 & 1.952 & 0.356 & 0.343 & 2.345 & 2.315 & 0.608\\
\cellcolor{gray!8}{\hspace{1em}5} & \cellcolor{gray!8}{100} & \cellcolor{gray!8}{$\infty$} & \cellcolor{gray!8}{0.310} & \cellcolor{gray!8}{1.954} & \cellcolor{gray!8}{1.935} & \cellcolor{gray!8}{0.351} & \cellcolor{gray!8}{0.320} & \cellcolor{gray!8}{2.324} & \cellcolor{gray!8}{2.301} & \cellcolor{gray!8}{0.608}\\
\addlinespace[0.4em]
\multicolumn{11}{l}{$p=5,\ n=500$}\\
\hspace{1em}5 & 500 & 1 & 0.127 & 1.341 & 1.335 & 0.143 & 0.145 & 2.322 & 2.315 & 0.678\\
\cellcolor{gray!8}{\hspace{1em}5} & \cellcolor{gray!8}{500} & \cellcolor{gray!8}{1.5} & \cellcolor{gray!8}{0.124} & \cellcolor{gray!8}{1.340} & \cellcolor{gray!8}{1.334} & \cellcolor{gray!8}{0.143} & \cellcolor{gray!8}{0.143} & \cellcolor{gray!8}{2.314} & \cellcolor{gray!8}{2.307} & \cellcolor{gray!8}{0.676}\\
\hspace{1em}5 & 500 & $\infty$ & 0.116 & 1.324 & 1.319 & 0.142 & 0.132 & 2.286 & 2.281 & 0.685\\
\addlinespace[0.4em]
\multicolumn{11}{l}{$p=20,\ n=100$}\\
\cellcolor{gray!8}{\hspace{1em}20} & \cellcolor{gray!8}{100} & \cellcolor{gray!8}{1} & \cellcolor{gray!8}{0.356} & \cellcolor{gray!8}{2.098} & \cellcolor{gray!8}{2.076} & \cellcolor{gray!8}{0.386} & \cellcolor{gray!8}{0.659} & \cellcolor{gray!8}{15.866} & \cellcolor{gray!8}{4.330} & \cellcolor{gray!8}{0.596}\\
\hspace{1em}20 & 100 & 1.5 & 0.345 & 2.091 & 2.070 & 0.384 & 0.640 & 15.853 & 4.312 & 0.596\\
\cellcolor{gray!8}{\hspace{1em}20} & \cellcolor{gray!8}{100} & \cellcolor{gray!8}{$\infty$} & \cellcolor{gray!8}{0.319} & \cellcolor{gray!8}{2.070} & \cellcolor{gray!8}{2.053} & \cellcolor{gray!8}{0.386} & \cellcolor{gray!8}{0.615} & \cellcolor{gray!8}{14.979} & \cellcolor{gray!8}{4.336} & \cellcolor{gray!8}{0.588}\\
\addlinespace[0.4em]
\multicolumn{11}{l}{$p=20,\ n=500$}\\
\hspace{1em}20 & 500 & 1 & 0.143 & 1.737 & 1.731 & 0.258 & 0.153 & 2.353 & 2.346 & 0.638\\
\cellcolor{gray!8}{\hspace{1em}20} & \cellcolor{gray!8}{500} & \cellcolor{gray!8}{1.5} & \cellcolor{gray!8}{0.142} & \cellcolor{gray!8}{1.732} & \cellcolor{gray!8}{1.726} & \cellcolor{gray!8}{0.257} & \cellcolor{gray!8}{0.151} & \cellcolor{gray!8}{2.345} & \cellcolor{gray!8}{2.339} & \cellcolor{gray!8}{0.637}\\
\hspace{1em}20 & 500 & $\infty$ & 0.131 & 1.712 & 1.707 & 0.255 & 0.138 & 2.323 & 2.318 & 0.641\\
\bottomrule
\end{tabular}
\begin{tablenotes}
\footnotesize
\item Bias, pooled RMSE, mean replicate-specific PEHE, and mean Regret summarize the Monte Carlo replicates as defined in the text.
BART denotes Bayesian additive regression trees; linear regression denotes the
parametric benchmark. The setting $\omega=\infty$ denotes the
monotonicity-limit construction.
\end{tablenotes}
\end{threeparttable}
\endgroup{}
\end{table}

Table~\ref{tab:sim1} summarizes the performance of BART and the parametric
linear model for estimating the CSACE surface. Across all combinations of
$n$, $p$, and $\omega$, both estimators had relatively small bias, suggesting
that when the proposed identification framework is correctly specified, CSACE
estimation is not primarily driven by systematic bias, even without imposing
monotonicity. Bias decreased as the sample size increased from $n=100$ to
$n=500$, with both methods achieving biases close to 0.1--0.15 in the larger
sample settings.

The two estimators differed more substantially in accuracy and stability. BART
consistently achieved smaller RMSE and PEHE than the linear model, indicating
more accurate recovery of the heterogeneous CSACE surface. The advantage of
BART was especially pronounced in the more challenging high-dimensional,
small-sample setting with $p=20$ and $n=100$, where the linear model exhibited
large RMSE values, reflecting occasional unstable replicates with extreme
estimation errors. The PEHE values for the linear model were also larger,
although less dramatically, consistent with PEHE being less sensitive than
Monte Carlo RMSE to rare extreme replicates.

BART also produced markedly smaller regret across all scenarios, indicating
fewer magnitude-weighted errors in the estimated direction of the conditional
treatment effect. This improvement is important because the sign of
$\mu_{11}(\bm X_i)$ determines whether treatment is estimated to be beneficial
or harmful for an individual in the always-survivor stratum. The results were
stable across $\omega=1$, $\omega=1.5$, and the monotonicity-limit setting
$\omega=\infty$, suggesting that the proposed estimator performs consistently
across different principal-stratum dependence structures. Overall, Experiment
1 shows that flexible BART outcome modeling improves recovery of the nonlinear
CSACE surface relative to a parametric linear benchmark, particularly when the
covariate dimension is larger and the sample size is limited.

\subsection{Experiment 2: Estimation of the CSACE-Based Variable Importance Measure}
\label{sec:sim-vim}

The second experiment evaluated estimation of the proposed CSACE-based
variable importance measure (VIM) within the estimated always-survivor stratum. We considered sample sizes
$n\in\{100,500\}$, covariate dimensions $p\in\{5,20\}$, and fixed
$\omega=1$, corresponding to conditional independence of the two potential
survival indicators. For each simulated data set, VIMs were estimated within
$\widehat{\mathcal A}$ using either BART or the parametric linear model. 

For benchmarking, we constructed an oracle VIM for each realized data set by
applying the same VIM algorithm to the true CSACE values
$\{\mu_{11}(\bm X_i):i\in\widehat{\mathcal A}\}$. Thus, the oracle target is a
data-set-level benchmark rather than a population-level analytic quantity.
This design isolates the effect of CSACE estimation error on VIM recovery,
while holding fixed the realized covariate distribution and the VIM
computation procedure. Performance was summarized by absolute bias, mean
squared error (MSE), and top-$k$ rank agreement between the estimated and
oracle VIM rankings. Rank agreement was measured using Kendall's $\tau$ for
$k\ge2$, with $k=1$ interpreted as exact recovery of the top-ranked variable.

Supplementary Figure~1 summarizes the numerical accuracy of VIM
estimation. Across all four $(p,n)$ settings, BART yielded smaller bias and
MSE than the parametric linear model. The improvement was especially clear
when $n=500$, where the linear model showed substantial bias and variability,
whereas BART remained concentrated closer to zero. When $p=20$, BART also
remained stable after the inclusion of additional noise covariates, suggesting
robustness to moderate-dimensional nuisance variation. These results indicate
that flexible estimation is important for accurately recovering the VIM when
CSACE heterogeneity is nonlinear and interaction-driven. Supplementary Figure~2 further evaluates whether the estimated VIM can
recover the oracle ordering of important variables. This ranking criterion is
directly relevant to the intended use of the VIM, because in real
applications the VIM is primarily used to identify a small set of
effect-modifying variables for interpretation and downstream subgroup
discovery. BART showed near-perfect recovery of the most important variables.
For $k=1$, BART recovered the top-ranked variable in all simulation
replicates. For $k=2$, BART again achieved essentially perfect agreement,
correctly identifying the two most important variables and their relative
ordering. As $k$ increased, the ranking task became more difficult because
moderately important variables were included, but BART continued to yield
substantially higher rank agreement than the parametric linear model.

In contrast, the linear model showed unstable ranking performance, with
Kendall's $\tau$ often near zero or negative. This indicates weak, and in
some cases reversed, agreement with the oracle VIM ordering. In summary,
Supplementary Figures~1 and~2 show that
flexible nuisance estimation improves not only the numerical accuracy of VIM
estimation but also its practical utility as a screening tool for identifying
interpretable sources of CSACE heterogeneity.

\section{Application to the ARDS Network Trial}
\subsection{Study Setting and Application Overview}
We apply the proposed method to the ARDS Network trial comparing lower and higher positive end-expiratory pressure (PEEP) strategies in patients with acute lung injury or acute respiratory distress syndrome~\citep{national2004higher}. As described in Section~\ref{sec1}, the trial provides a useful setting for evaluating conditional survivor causal effects because the original population-level comparisons showed no significant improvement in major clinical outcomes under the higher-PEEP strategy, while the clinical mechanisms of PEEP suggest plausible benefit--harm heterogeneity across patients~\citep{sahetya2017fifty}. In addition, comparing two active ventilation strategies makes monotonicity questionable~\citep{tong2025semiparametric}, motivating the use of the non-monotonic CSACE framework developed in this paper.

For our analysis, let \(Z_i\in\{0,1\}\) denote randomized treatment assignment, where \(Z_i=0\) indicates the lower-PEEP strategy and \(Z_i=1\) indicates the higher-PEEP strategy. Let \(D_i\in\{0,1\}\) denote the post-treatment survival indicator by 60 days, with \(D_i=1\) if patient \(i\) survives to 60 days and \(D_i=0\) otherwise. Let \(X_i\) denote baseline covariates measured before randomization. The non-mortality outcome \(Y_i\) is defined as the time to discharge home, censored by the date of last contact when applicable. Because discharge time is only meaningfully defined among patients who survive long enough for post-baseline follow-up to be observed, this outcome is subject to truncation by death. Thus, survivor causal effects provide a more appropriate target than an unrestricted population-level contrast of discharge-related outcomes.

Our analysis has two goals. First, we estimate CSACEs for the post-survival outcome without imposing monotonicity, using the proposed sensitivity framework to allow for the possibility that some patients would survive under lower PEEP but not under higher PEEP. Second, we use the posterior distribution of individualized CSACEs to assess treatment effect heterogeneity, identify important baseline covariates, and construct clinically interpretable subgroups with evidence of benefit or harm from the higher-PEEP strategy. This analysis illustrates how the proposed framework can move beyond the original average comparisons by connecting survivor causal estimands, Bayesian heterogeneity estimation, and subgroup discovery in a trial where death truncation and non-monotonicity are central concerns.

\subsection{Identifying Likely Always-Survivors}

Following the practical strategy in previous work \citep{chen2024bayesian}, we first identify a tangible subset of observed survivors with higher posterior probabilities of belonging to the always-survivor stratum before examining CSACE heterogeneity. For each individual with $D_i=1$, let $\hat\gamma_i$ denote the posterior mean of $\Pr\{D_i(0)=1,D_i(1)=1\mid D_i=1,\bm X_i\}$, as defined in Section~\ref{sec2}. Under the anchor sensitivity value $\xi=0.25$, we classify individual $i$ as a likely always-survivor if and only if
\[
\hat\gamma_i>\hat{\bar\gamma}=0.695,
\]
where $\hat{\bar\gamma}$ is the estimated marginal probability of always-survivor membership among observed survivors. Thus, 69.5\% is a data-adaptive cutoff, and the rule selects observed survivors whose posterior mean membership probability exceeds the estimated marginal benchmark.

Our rule shares the objective of separating uncertainty about principal-stratum membership from subsequent exploration of response heterogeneity \citep{chen2024bayesian}, but it differs in two important respects. Under the monotonicity assumption used in that work, survivors in one treatment arm are known always-survivors, and a posterior-probability threshold is applied only to participants with ambiguous membership; the threshold $p=0.8$ was calibrated so that the size of the resulting likely always-survivor subset matched the posterior mean marginal prevalence of the stratum. Without monotonicity, no treatment-arm-specific group of survivors is deterministically known to belong to the always-survivor stratum. We therefore apply a common posterior-mean membership rule to all observed survivors and use the estimated marginal membership probability as the cutoff. This construction preserves the practical value of a well-defined subgroup for heterogeneity analysis while explicitly retaining uncertainty induced by non-monotonic principal-stratum membership.

\subsection{Choice of Sensitivity Parameters}
As discussed in Section~\ref{sec2}, to implement the sensitivity analysis, we parameterized the association between the two potential survival indicators through the log-odds ratio parameter $\xi$, where $\exp(\xi)$ is the conditional odds ratio measuring the association between survival under lower PEEP and survival under higher PEEP, given baseline covariates. Larger values of $\xi$ correspond to stronger positive dependence between $D(0)$ and $D(1)$, and therefore to a larger estimated always-survivor stratum. The value $\xi=0$ represents conditional independence between the two potential survival indicators, while positive values encode the clinically plausible idea that patients with better baseline survival prognosis under one ventilation strategy are also more likely to survive under the other. We varied $\xi$ over the range $[-0.5,1.5]$, corresponding to conditional odds ratios from approximately $0.61$ to $4.48$. This range was chosen to cover a broad but interpretable spectrum of principal-stratum dependence. The lower end allows mild negative dependence, thereby explicitly permitting non-monotone survival behavior and avoiding an implicit assumption that the two survival potentials must be positively associated. The middle of the range includes conditional independence and weak positive dependence. The upper end allows a substantially stronger shared-survival tendency without imposing the deterministic monotonicity assumption that would rule out patients harmed by higher PEEP. 

\begin{figure}[htpb!] 
\centering 
\includegraphics[width=0.88\textwidth]{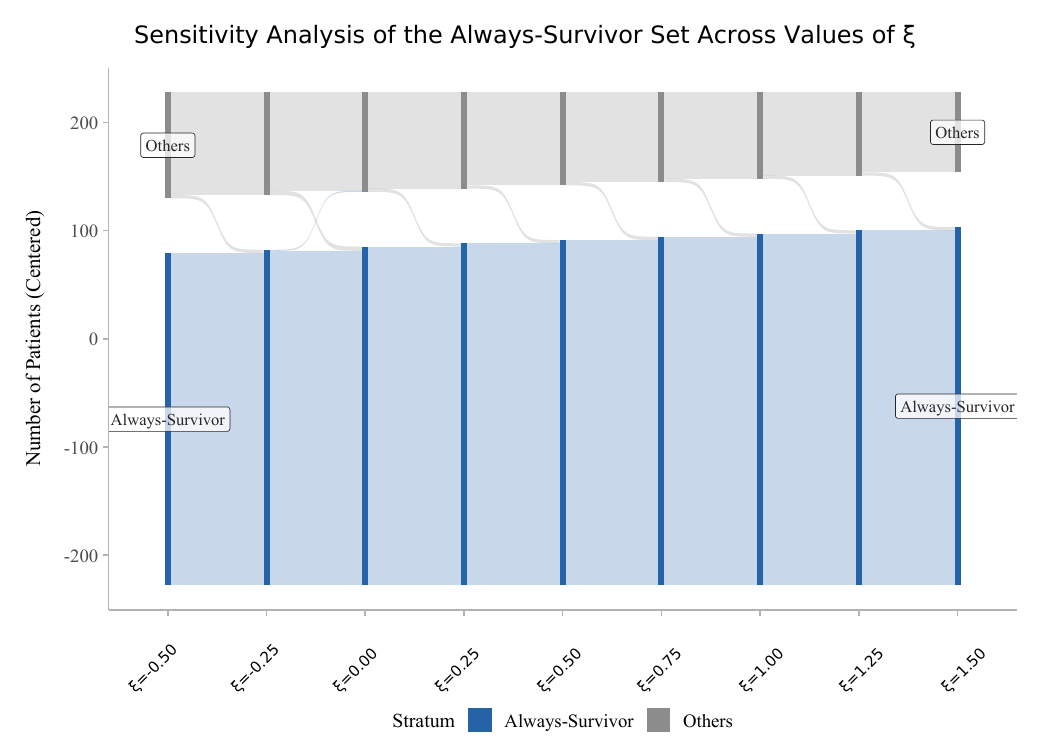} 
\caption{ Sensitivity analysis of the estimated always-survivor stratum across values of $\xi$. Blue bars represent patients classified as likely always-survivors, and gray bars represent all other observed survivors. As $\xi$ increases, stronger positive dependence between $D(0)$ and $D(1)$ leads to a larger estimated always-survivor stratum. } \label{fig:realdata_sensitivity_as} 
\end{figure} 

Figure~\ref{fig:realdata_sensitivity_as} shows how the estimated always-survivor set changes across the sensitivity range. As expected, the number of patients classified as always-survivors increases with $\xi$: stronger positive association between survival under lower and higher PEEP implies that a larger fraction of observed survivors are likely to belong to the stratum who would survive under both treatment strategies. Importantly, the change is gradual rather than abrupt. This indicates that the estimated principal-stratum membership is not driven by a sharp instability at a particular sensitivity value, but instead evolves smoothly as the assumed cross-world survival dependence is strengthened. For the remaining analysis, we use $\xi=0.25$ as the anchor value. This choice corresponds to a conditional odds ratio of $\exp(0.25)\approx 1.28$, representing weak positive dependence between $D(0)$ and $D(1)$. Clinically, this reflects the reasonable expectation that survival prognosis is partly shared across the two active ventilation strategies, because baseline physiologic severity and respiratory mechanics are prognostic for outcomes in mechanically ventilated patients with ARDS~\citep{amato2015driving}. Thus, patients who would survive under lower PEEP may be somewhat more likely to survive under higher PEEP as well, even though the possibility of strategy-specific harm remains. At the same time, $\xi=0.25$ remains close to conditional independence and far from a strong monotonicity-like assumption. It therefore provides a conservative anchor for the primary analysis while preserving the central feature of our framework: the possibility that some patients may be harmed by the higher-PEEP strategy.

\subsection{Estimation of CSACE, VIM and Subgroup Discovery}
\begin{figure}[htpb!]
\centering
\includegraphics[width=0.95\textwidth]{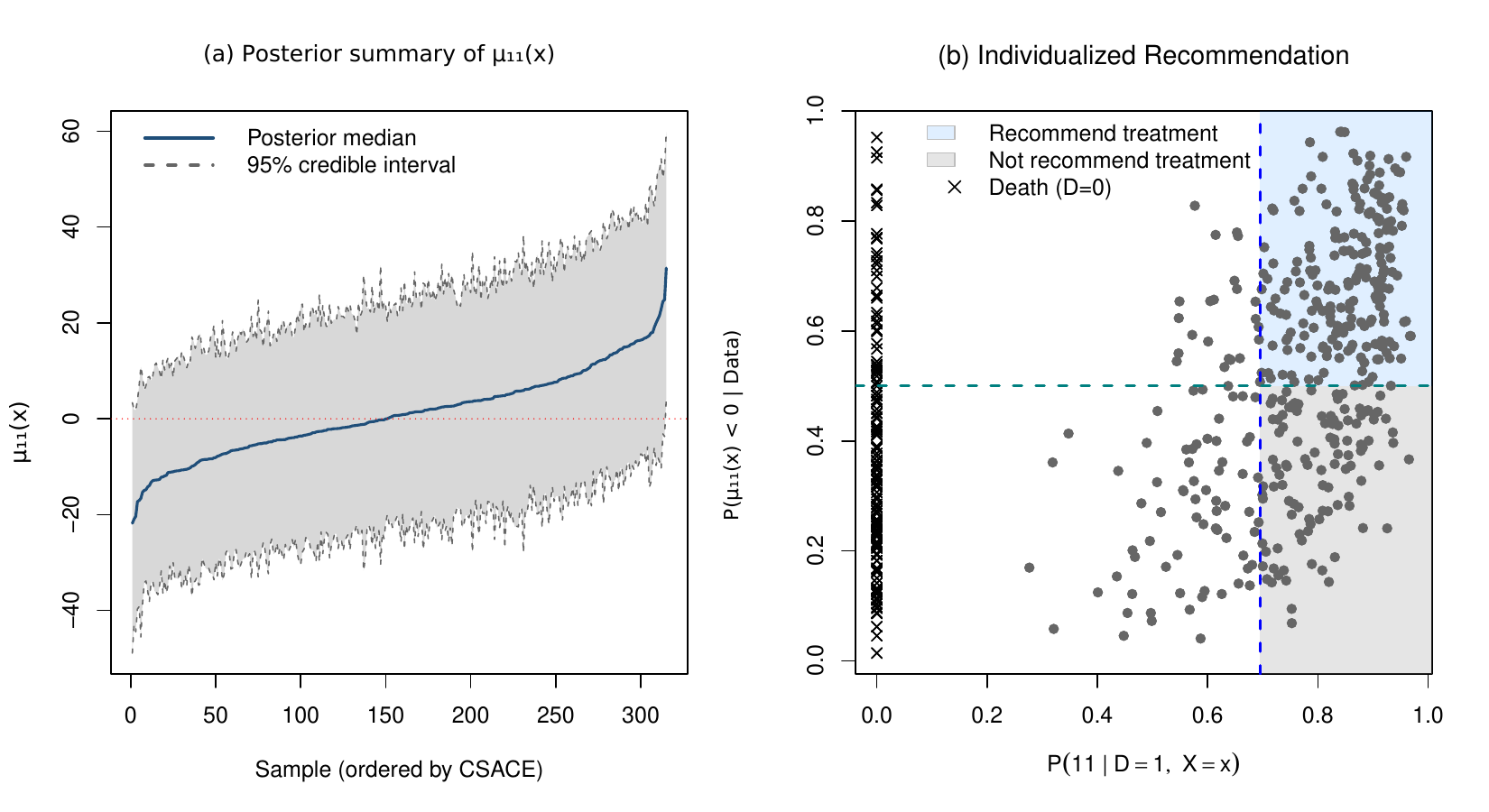}
\caption{
Estimated individualized CSACE and exploratory treatment classification under the anchor
sensitivity value $\xi=0.25$.
(a) Posterior summary of the CSACE $\mu_{11}(\bm x)$ within estimated
always-survivors. The solid curve shows the posterior median, and the dashed
curves show the 95\% credible interval, with patients ordered by posterior
median CSACE.
(b) Exploratory individualized treatment classification. The x-axis is the
individual posterior probability of being an always-survivor,
$\Pr\{D(0)=1,D(1)=1\mid D=1,\bm X=\bm x\}$, and the y-axis is the posterior
probability of treatment benefit,
$\Pr\{\mu_{11}(\bm x)<0\mid \text{Data}\}$. Patients with posterior
always-survivor probability greater than 69.5\% are classified as likely
always-survivors. Among them, patients with
$\Pr\{\mu_{11}(\bm x)<0\mid \text{Data}\}>0.5$ are more likely to benefit from
higher PEEP. Therefore, the upper-right blue shaded region is classified as
favoring higher PEEP, whereas the lower-right gray shaded region is classified
as not favoring higher PEEP.
Patients with $D=0$ are shown separately as crosses.
}
\label{fig:real_csace}
\end{figure}

Figure~\ref{fig:real_csace} summarizes the posterior distribution of
individualized CSACEs under the proposed non-monotonic principal-stratification
framework. Panel (a) shows the posterior median and 95\% credible interval of
$\mu_{11}(\bm x)$ among estimated always-survivors, ordered by posterior
median CSACE. The posterior medians span both negative and positive values,
indicating substantial heterogeneity in the survivor causal effect of higher
PEEP on time to discharge home. Because negative CSACE values correspond to
shorter discharge time under higher PEEP, the left portion of the curve
represents patients more likely to benefit from higher PEEP, whereas the right
portion represents patients for whom higher PEEP may delay discharge.

Panel (b) translates these posterior summaries into an exploratory
classification display by combining two distinct sources of uncertainty:
uncertainty about always-survivor membership and uncertainty about the
direction of the CSACE. Patients with
\[
\Pr\{D(0)=1,D(1)=1\mid D=1,\bm X=\bm x\}>0.695
\]
were classified as likely always-survivors. Within this estimated
always-survivor group, patients with
\[
\Pr\{\mu_{11}(\bm x)<0\mid \text{Data}\}>0.5
\]
were classified as more likely to benefit from higher PEEP. This display
illustrates how the proposed framework
separates principal-stratum uncertainty from treatment-effect uncertainty:
the x-axis determines whether the survivor causal estimand is considered
relevant for a patient, while the y-axis determines whether the posterior
CSACE distribution favors benefit or harm. In this sense, the classification
rule is not based solely on a point estimate of treatment effect, but instead
combines estimated always-survivor membership with posterior evidence for a
negative CSACE.

We next examined which baseline covariates contributed most to heterogeneity
in the estimated CSACE surface. Applying the proposed CSACE-based VIM within
the estimated always-survivor stratum identified five leading pre-treatment covariates: age at enrollment,
sex, height, plateau pressure (PSTATVC), and arterial blood pH (ARTPH). For DaCIT, we retained the 10 highest-ranked covariates; the main text focuses on the leading five, while the complete ranking is reported in the Supplementary Material. Total respiratory rate (TRESPR), although not among the five emphasized here, was therefore included in the active set available to the tree.
Although these variables should not be interpreted as externally validated
effect modifiers, they align with clinically meaningful domains of ARDS
severity, ventilatory mechanics, and lung-protective ventilation
\citep{acute2000ventilation,amato2015driving,griffiths2019guidelines}.

Age may capture differences in physiologic reserve, comorbidity burden,
recovery trajectory, and tolerance of the hemodynamic consequences of higher
PEEP. Sex and height are directly connected to lung-protective ventilation
because tidal-volume targets in ARDS are based on predicted body weight, which
is calculated from sex and height. Recent work has questioned whether these
formulas generalize equally across patient groups and whether lung-protective
ventilation is sex neutral
\citep{acute2000ventilation,mcnicholas2019demographics,sarma2025evaluating,chiumello2026lung}.
Even in a randomized trial of PEEP assignment, delivered mechanical exposure
may therefore differ with body size and sex. Such differences could contribute
to the observed sex-related heterogeneity, although the present analysis does
not establish this mechanism.

Plateau pressure reflects respiratory-system mechanics and the balance between
alveolar recruitment and over-distention. A higher baseline plateau pressure
may indicate less room to increase end-inspiratory lung volume without
over-distention; when recruitability is limited, additional PEEP may
preferentially distend already aerated lung
\citep{sahetya2017fifty,luecke2005clinical}. Arterial pH summarizes acid--base
derangement and systemic physiologic severity. A lower pH may indicate shock
or severe respiratory or metabolic compromise, potentially identifying
patients who are less able to tolerate reductions in venous return or other
hemodynamic effects of higher PEEP
\citep{knaus1991apache,xingzheng2024impact}. These interpretations provide
clinical context for the VIM ranking but are not direct tests of the underlying
mechanisms.

The VIM results suggest that heterogeneity in the survivor
effect of higher PEEP is concentrated along clinically interpretable dimensions of baseline vulnerability, body-size-related ventilation targets, respiratory mechanics, and acute physiologic severity. The VIM ranking was therefore used to define the candidate effect-modifier set for the subsequent distribution-aware subgroup analysis.

Finally, we used the posterior individualized CSACE distributions to construct
a distribution-aware conditional inference tree among the estimated
always-survivors. This analysis aims to translate the posterior heterogeneity
shown in Figure~\ref{fig:real_csace} into clinically interpretable subgroups
with distinct benefit--harm profiles.

\begin{figure}[htpb!]
\centering
\includegraphics[width=0.98\textwidth]{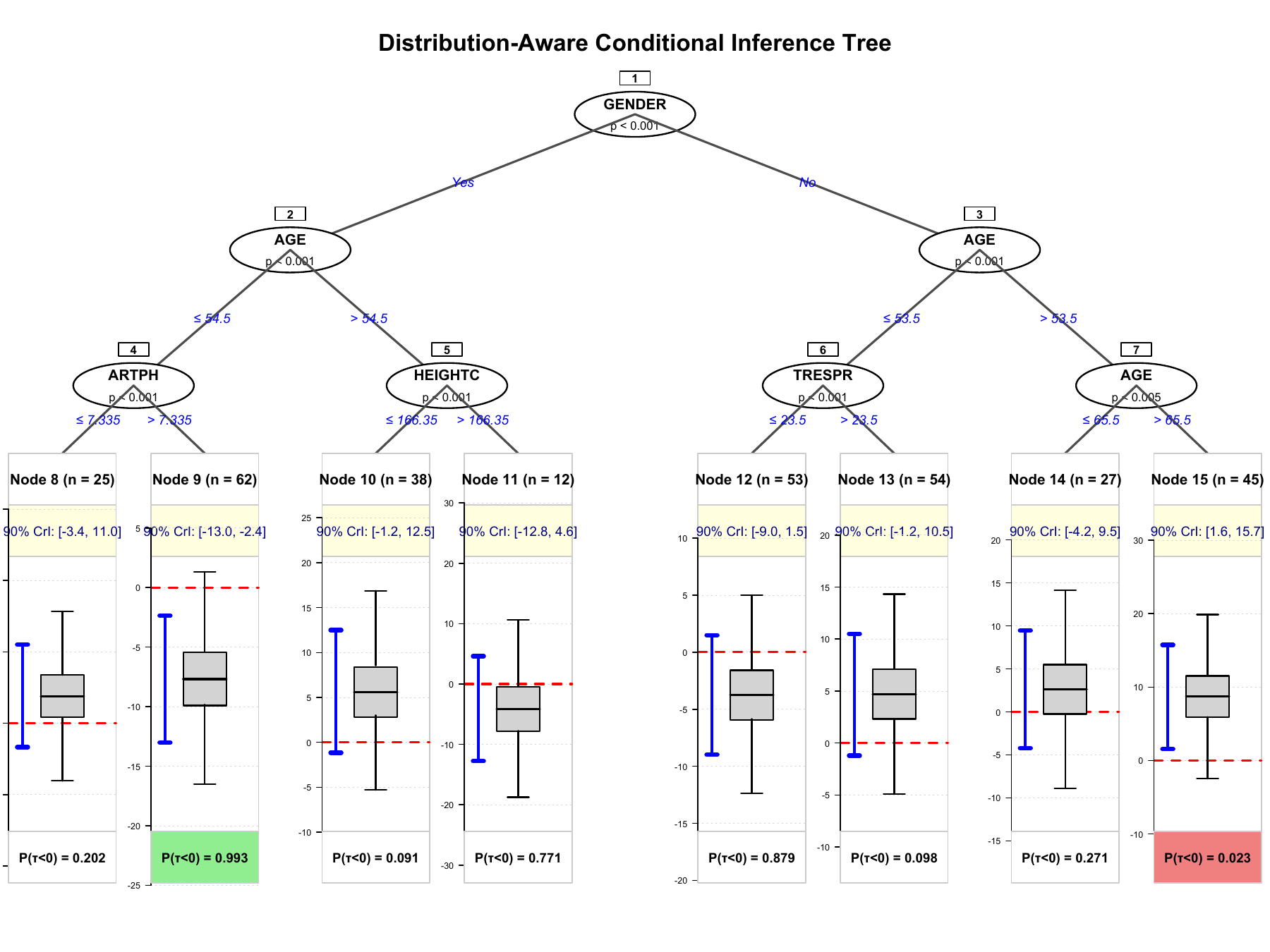}
\caption{
Distribution-aware conditional inference tree based on posterior
individualized CSACE distributions among estimated always-survivors. Internal
nodes show the selected splitting variables and their permutation-based
$p$-values. Terminal nodes summarize the posterior distribution of
$\mu_{11}(\bm x)$ within each subgroup, including the 90\% credible interval
and the posterior probability of benefit,
$\Pr\{\mu_{11}(\bm x)<0\}$. Negative CSACE values correspond to shorter time
to discharge home under higher PEEP. Green shading indicates a subgroup with
strong posterior evidence of benefit from higher PEEP, whereas red shading
indicates a subgroup with strong posterior evidence of harm.
}
\label{fig:real_tree}
\end{figure}

Figure~\ref{fig:real_tree} presents the resulting distribution-aware
conditional inference tree. The first split was by sex, indicating that
the posterior CSACE distributions differed most strongly between female and
male patients. Subsequent splits involved age, arterial pH, height, respiratory
rate, and age again within the male branch. The numerical cutpoints should be
interpreted as data-adaptive thresholds selected by the DaCIT algorithm rather
than as externally validated clinical decision thresholds. Nevertheless, the
selected variables and approximate cutpoint regions are clinically
interpretable. Sex and height determine predicted body weight and therefore
the target tidal volume used in lung-protective ventilation, and prior studies
have reported that shorter or female patients may receive relatively higher
tidal volumes when PBW-based ventilation is not carefully implemented
\citep{acute2000ventilation,linares2015standardizing,sasko2018size,self2023disparities}.
Age is a well-established prognostic factor in ARDS, with older patients
showing worse outcomes in multiple studies, including analyses using age
thresholds around 65 years \citep{brown2020impact,kao2018survival,schouten2019increased}.
Arterial pH and respiratory rate are markers of acute physiologic derangement,
and arterial pH has been associated with survival in ARDS analyses
\citep{amato2015driving,knaus1991apache}. This structure is therefore
consistent with the VIM results, which identified demographic characteristics,
ventilatory mechanics, and physiologic severity markers as important sources
of CSACE heterogeneity.

The most pronounced benefit subgroup was Node 9, consisting of female patients
aged $\leq 54.5$ years with arterial pH greater than 7.335. In this subgroup,
the 90\% credible interval was entirely below zero, $[-13.0,-2.4]$, and the
posterior probability of benefit was
$\Pr\{\mu_{11}(\bm x)<0\}=0.993$. Because negative CSACE values indicate
shorter time to discharge home under higher PEEP, this node provides the
strongest evidence that higher PEEP may accelerate post-survival recovery
among estimated always-survivors. Clinically, this pattern is plausible but
exploratory: younger patients and patients with more preserved acid--base
status may have greater physiologic reserve, while sex- and height-related
differences in predicted body weight may influence the mechanical consequences
of a higher-PEEP strategy \citep{acute2000ventilation,mcnicholas2019demographics,amato2015driving}.
In such patients, the potential benefits of recruitment and oxygenation under
higher PEEP may outweigh the risks of over-distension or hemodynamic compromise
\citep{sahetya2017fifty,luecke2005clinical}.

In contrast, the clearest potential harm subgroup was Node 15, consisting of
older male patients aged $>65.5$ years. This subgroup had a 90\% credible
interval entirely above zero, $[1.6,15.7]$, and a very low posterior
probability of benefit,
$\Pr\{\mu_{11}(\bm x)<0\}=0.023$. This finding is clinically coherent because
older patients with ARDS have higher baseline vulnerability and worse outcomes,
and the age cutpoint selected by the tree lies near age ranges commonly used to
define older ARDS populations \citep{brown2020impact,kao2018survival,schouten2019increased}.
Higher PEEP can improve oxygenation, but it may also increase over-distension
and alter cardiopulmonary mechanics, including pulmonary vascular resistance,
right ventricular loading, venous return, and cardiac output; such effects may
be less well tolerated in vulnerable patients \citep{sahetya2017fifty,luecke2005clinical,xingzheng2024impact}.

Other terminal nodes showed more uncertain or intermediate effects, with
credible intervals crossing zero. For example, Node 12, representing younger
male patients with respiratory rate $\leq 23.5$, showed suggestive evidence of
benefit but with a credible interval crossing zero, whereas Nodes 8, 10, 13,
and 14 showed weaker or more ambiguous posterior evidence. Taken together, the
tree suggests that the benefit--harm profile of higher PEEP is not uniform
among patients who would survive under either treatment strategy. Instead, the
survivor effect appears to vary along clinically meaningful dimensions related
to sex, age, respiratory physiology, body size, and baseline severity.
These results should be interpreted as exploratory subgroup findings, but they
illustrate how the proposed framework can translate posterior CSACE
heterogeneity into clinically interpretable benefit and harm profiles without
relying on monotonicity.

\begin{table}[htpb!]
\centering
\footnotesize
\begin{threeparttable}
\caption{Hypothesis-generating clinical framework for interpreting heterogeneity in response to higher PEEP.}
\label{tab:application_summary}
\begin{tabular}{>{\raggedright\arraybackslash}p{0.15\textwidth}>{\raggedright\arraybackslash}p{0.50\textwidth}>{\raggedright\arraybackslash}p{0.25\textwidth}}
\toprule
Analysis signal & Literature-supported clinical interpretation & Limits of interpretation \\
\midrule
Age & Older age may serve as a proxy for lower physiologic reserve, reduced capacity for recovery, and greater vulnerability to cardiopulmonary stress; age is also prognostic in ARDS cohorts~\citep{brown2020impact,kao2018survival,schouten2019increased}. & Age was measured at baseline, but physiologic reserve was not directly measured. \\
\addlinespace
Sex and height & Sex and height jointly determine predicted body weight (PBW). PBW-based formulas may yield different relative ventilatory exposures among shorter, female, and older patients~\citep{sarma2025evaluating,chiumello2026lung}. & Sex and height were selected, but formula-related PBW miscalibration and individual lung size were not evaluated. \\
\addlinespace
Plateau pressure (PSTATVC) & Baseline plateau pressure reflects respiratory-system mechanics and the available pressure margin. When recruitability is limited, additional PEEP may preferentially distend already aerated lung~\citep{sahetya2017fifty,gattinoni2006lung,constantin2010lung}. & Plateau pressure was measured, but lung recruitability and regional morphology were not. \\
\addlinespace
Arterial pH (ARTPH) & Low pH marks acute respiratory or metabolic derangement and may identify patients with limited tolerance of reductions in venous return or other hemodynamic effects of higher PEEP~\citep{knaus1991apache,luecke2005clinical,xingzheng2024impact}. & pH was measured, but its cause and the acute hemodynamic response to PEEP were not. \\
\bottomrule
\end{tabular}
\begin{tablenotes}
\footnotesize
\item[] These interpretations use prior clinical and physiologic literature to contextualize variables prioritized by the CSACE-VIM and DaCIT analyses. The proposed pathways were not directly tested in this study; the identified variables are not validated effect modifiers or treatment-selection rules.
\end{tablenotes}
\end{threeparttable}
\end{table}

Table~\ref{tab:application_summary} links the leading variables to literature-supported clinical hypotheses while distinguishing these interpretations from mechanisms not measured in the trial. The complete CSACE-VIM ranking and subgroup results across all values of $\xi$ are reported in the Supplementary Material.

\section{Discussion}
We developed a monotonicity-free framework for heterogeneous survivor causal effects that retains both protected and harmed principal strata. BART provides individualized CSACE posterior draws, the CSACE-VIM screens covariates within the estimated always-survivor stratum, and DaCIT uses the full posterior distributions to construct interpretable subgroups. Simulations showed that BART improved nonlinear CSACE estimation, directional decisions, and recovery of important variables relative to linear modeling.

In the ARDS trial, the average survivor effect of higher PEEP on time to discharge home was not clearly different from zero~\citep{tong2025semiparametric}, but individualized effects varied substantially. The strongest benefit profile comprised younger female patients with relatively preserved arterial pH, whereas the clearest potential-harm profile comprised older male patients. These findings are biologically plausible in light of variation in physiologic reserve, PBW-based ventilation, recruitment, over-distention, and hemodynamic vulnerability, but the data do not establish these mechanisms. The subgroups and data-adaptive cutpoints are exploratory and require external validation before clinical use.

The framework is particularly relevant to critical care trials in which death truncates recovery outcomes and active treatments have competing physiologic effects. It provides a transparent route from an inconclusive average result to testable hypotheses about benefit--harm heterogeneity while preserving uncertainty in individualized effects. Two limitations concern identification. First, always-survivor membership depends on a sensitivity parameter governing the unobserved association between $D(0)$ and $D(1)$. The parameter cannot be learned from observed data, but its log-odds-ratio interpretation permits conclusions to be examined across clinically plausible values rather than ruling out the harmed stratum. Second, just like unconfoundedness in observational study contexts, conditional principal ignorability remains empirically unverifiable and requires adequate baseline prognostic adjustment. A useful extension would introduce a second sensitivity parameter for violations of principal ignorability, trading stronger transparency for additional clinical input when specifying plausible ranges.

\section{Significance Statement}

Clinical trials in critically ill patients often measure outcomes such as time to discharge that are meaningful only among patients who survive. Standard analyses can be misleading when treatment affects survival differently across patients. We develop a flexible method that estimates how treatment effects vary among patients who would survive under either treatment strategy, without assuming that treatment cannot harm survival. The method also identifies baseline characteristics associated with benefit or harm while retaining uncertainty in individualized estimates. Applied to a trial of higher versus lower positive end-expiratory pressure for acute respiratory distress syndrome, the approach suggests clinically interpretable variation hidden by the trial's null average result and generates hypotheses for future validation.

\bibliographystyle{imsart-nameyear} 
\bibliography{bibliography.bib}       


\end{document}


\begin{frontmatter}
\title{Supplementary Material to ``Heterogeneous survivor average causal effects beyond monotonicity: Applications to a clinical trial evaluating mechanical ventilation strategies"}
\runtitle{SGL}

\begin{aug}
\begin{aug}
\author[A]{\fnms{Zihan}~\snm{Zhu}\ead[label=e1]{zihan.zhu@yale.edu}},
\author[A]{\fnms{Guangyu}~\snm{Tong}\ead[label=e2]{guangyu.tong@yale.edu}},
\author[B]{\fnms{Fernando Godinho}~\snm{Zampieri}\ead[label=e3]{fzampier@ualberta.ca}},
\author[D]{\fnms{Snigdha}~\snm{Jain}\ead[label=e4]{Snigdha.Jain@yale.edu}},
\author[C]{\fnms{Michael O.}~\snm{Harhay}\ead[label=e5]{mharhay@pennmedicine.upenn.edu}},
\author[A]{\fnms{Fan}~\snm{Li}\ead[label=e6]{fan.f.li@yale.edu}},

\address[A]{Department of Biostatistics, Yale School of Public Health%
  \printead[presep={,\ }]{e1,e2,e6}}
\address[B]{Department of Critical Care, University of Alberta%
  \printead[presep={,\ }]{e3}}
\address[C]{Department of Biostatistics, Epidemiology, and Informatics,
  University of Pennsylvania\printead[presep={,\ }]{e5}}
\address[D]{Section of Pulmonary, Critical Care, and Sleep Medicine,
  Department of Internal Medicine, Yale University School of Medicine%
  \printead[presep={,\ }]{e4}}
\end{aug}

\address[A]{Department of Biostatistics, Yale School of Public Health\printead[presep={,\ }]{e1}}
\address[A]{Department of Biostatistics, Yale School of Public Health\printead[presep={,\ }]{e2}}

\end{aug}

\end{frontmatter}



\section{Proof of Proposition 1}

\begin{proof}
We derive the identification of each component in turn. By SUTVA and Assumption 1,
\begin{equation*}
\begin{aligned}
m_{11}(\bm x)
&= \mathbb{E}[Y_i \mid Z_i=1,\, D_i=1,\, \bm X_i=\bm x] \\
&= \mathbb{E}[Y_i(1) \mid D_i(1)=1,\, \bm X_i=\bm x].
\end{aligned}
\end{equation*}
Applying the law of total expectation over principal strata with $D_i(1)=1$, i.e., strata $(1,1)$ and $(0,1)$,
\begin{equation*}
\begin{aligned}
&\mathbb{E}[Y_i(1) \mid D_i(1)=1,\, \bm X_i=\bm x]
\\ =&  \mathbb{E}[Y_i(1) \mid D_i(0)=1,\, D_i(1)=1,\, \bm X_i=\bm x]
   \cdot \mathbb{P}(D_i(0)=1 \mid D_i(1)=1,\, \bm X_i=\bm x) \\
\quad +&  \mathbb{E}[Y_i(1) \mid D_i(0)=0,\, D_i(1)=1,\, \bm X_i=\bm x]
   \cdot \mathbb{P}(D_i(0)=0 \mid D_i(1)=1,\, \bm X_i=\bm x).
\end{aligned}
\end{equation*}
By Assumption 2 with $z=1$, both conditional expectations on the right-hand side equal $\mathbb{E}[Y_i(1) \mid D_i(1)=1,\, \bm X_i=\bm x]$, so the mixture collapses and
\begin{equation*}
m_{11}(\bm x)= \mathbb{E}[Y_i(1) \mid D_i(0)=1,\, D_i(1)=1,\, \bm X_i=\bm x].
\end{equation*}
By an analogous argument with $z=0$,
\begin{equation*}
\begin{aligned}
m_{01}(\bm x)
& = \mathbb{E}[Y_i \mid Z_i=0,\, D_i=1,\, \bm X_i=\bm x] \\
& = \mathbb{E}[Y_i(0) \mid D_i(0)=1,\, \bm X_i=\bm x] \\
& = \mathbb{E}[Y_i(0) \mid D_i(0)=1,\, D_i(1)=1,\, \bm X_i=\bm x].
\end{aligned}
\end{equation*}
Combining these results,
\begin{equation*}
\mu_{11}(\bm x)
=m_{11}(\bm x) - m_{01}(\bm x). 
\end{equation*}
\end{proof}

\subsection{Derivation of the always-survivor posterior probability}

Let $e_{11}(\bm x)=\mathbb P\{D(0)=1,D(1)=1\mid\bm X=\bm x\}$, $p_z(\bm x)=\mathbb P(D=1\mid Z=z,\bm X=\bm x)$, and $\pi(\bm x)=\mathbb P(Z=1\mid\bm X=\bm x)$. For an observed survivor, the law of total probability gives
\[
\gamma_i=\sum_{z\in\{0,1\}}
\mathbb P\{D_i(0)=1,D_i(1)=1\mid Z_i=z,D_i=1,\bm X_i=\bm x\}
\mathbb P(Z_i=z\mid D_i=1,\bm X_i=\bm x).
\]
Under treatment ignorability, the first probability in each summand is $e_{11}(\bm x)/p_z(\bm x)$. Bayes' theorem gives
\[
\mathbb P(Z_i=1\mid D_i=1,\bm X_i=\bm x)
=\frac{p_1(\bm x)\pi(\bm x)}{p_1(\bm x)\pi(\bm x)+p_0(\bm x)\{1-\pi(\bm x)\}},
\]
with the analogous expression for $Z_i=0$. Substitution yields
\[
\gamma_i=
\frac{e_{11}(\bm x)}{p_1(\bm x)\pi(\bm x)+p_0(\bm x)\{1-\pi(\bm x)\}}.
\]
The marginal always-survivor proportion among observed survivors is $\bar\gamma=\mathbb E(\gamma_i\mid D_i=1)$ and is estimated by the sample mean of $\hat\gamma_i$ among observed survivors.

\section{Detailed Introduction of Data Generating Processes}
\label{sec:dgp-csace-vim}

We considered data-generating processes designed to evaluate CSACE estimation
and variable-importance learning under nonlinear principal-stratum assignment,
heterogeneous potential outcomes, and noise covariates. For each sample size
$n$, we generated independent observations
\[
\{(\bm X_i,Z_i,D_i,Y_i)\}_{i=1}^n,
\qquad
\bm X_i=(X_{i1},\ldots,X_{ip})^\top .
\]
The first five covariates were allowed to affect the principal strata and
outcome surfaces, whereas the remaining covariates were pure noise variables
included as negative controls. For each $j=1,\ldots,p$,
\[
X_{ij}\stackrel{\mathrm{i.i.d.}}{\sim}\mathrm{TN}(-10,10;0,1),
\]
where $\mathrm{TN}(-10,10;0,1)$ denotes a standard normal distribution
truncated to the interval $[-10,10]$.

Treatment assignment was randomized as $Z_i\sim\mathrm{Bernoulli}(0.5)$ independently of $\bm X_i$, matching both the target trial setting and the randomized design of the motivating application.

Let
\[
(D_i(0),D_i(1))\in\{0,1\}^2
\]
denote the two potential intermediate variables under control and treatment.
Instead of specifying the four principal-stratum probabilities directly, we
first generated the marginal principal scores
\[
p_0(\bm X_i)=\Pr\{D_i(0)=1\mid \bm X_i\},
\qquad
p_1(\bm X_i)=\Pr\{D_i(1)=1\mid \bm X_i\}.
\]
Write
\[
x_j=X_{ij},\qquad j=1,\ldots,5,
\]
and define the smooth gate
\[
\sigma(t)=\{1+\exp(-t)\}^{-1}.
\]
We constructed several nonlinear gating functions:
\[
g_{1s}=\sigma\{2(x_1-0.1)\},
\qquad
g_{2s}=\sigma\{2(x_2+x_3-0.25)\},
\]
\[
g_{4s}=\sigma\{2(x_2-x_3)\},
\qquad
g_{5s}=\sigma\{2(-x_1-0.05)\},
\]
together with a smooth rectangular gate
\[
r_s=\sigma\{2(x_4+0.2)\}\sigma\{2(0.1-x_5)\},
\]
and a smooth XOR-type gate
\[
q_s=\sigma(2x_1)+\sigma(2x_2)-2\sigma(2x_1)\sigma(2x_2).
\]
The logit-scale marginal principal scores were
\[
\eta_0(\bm x)
=
-0.2
+0.8x_3^2
-0.7x_4x_5
+1.2g_{1s}
-1.0g_{2s}
+1.5g_{1s}\sigma\{2(x_4x_5-0.1)\}
-1.0q_s
+0.3\max(x_3,0),
\]
and
\[
\eta_1(\bm x)
=
-0.1
+0.6x_1x_3
+1.2g_{1s}
-1.4g_{4s}
+1.2g_{5s}
-1.6g_{2s}g_{4s}
+1.3r_s .
\]
We then set
\[
p_0(\bm x)=\mathrm{expit}\{c_\eta\eta_0(\bm x)\},
\qquad
c_\eta=0.75.
\]
To encourage empirical monotonicity, we shifted the treatment-side marginal
score upward when the control-side logit score was larger. Specifically, with
$\kappa=0.75$ and $b=0.25$, define
\[
\eta_1^{\mathrm{adj}}(\bm x)
=
\eta_1(\bm x)
+
b
+
\kappa\max\{\eta_0(\bm x)-\eta_1(\bm x),0\},
\]
and set
\[
p_1(\bm x)
=
\mathrm{expit}\{c_\eta\eta_1^{\mathrm{adj}}(\bm x)\}.
\]

Given $p_0(\bm x)$ and $p_1(\bm x)$, the four principal-stratum probabilities
were determined through the conditional odds ratio $\omega=\exp(\xi)$. In the
simulations, $\omega$ was allowed to be either a common scalar or a
subject-specific vector. Let
\[
e_{d_0d_1}(\bm x)
=
\Pr\{D(0)=d_0,D(1)=d_1\mid \bm X=\bm x\},
\qquad
(d_0,d_1)\in\{0,1\}^2.
\]
The stratum probabilities satisfy the marginal constraints
\[
e_{10}(\bm x)+e_{11}(\bm x)=p_0(\bm x),
\qquad
e_{01}(\bm x)+e_{11}(\bm x)=p_1(\bm x),
\]
and the odds-ratio constraint
\[
\omega
=
\frac{e_{11}(\bm x)e_{00}(\bm x)}
     {e_{10}(\bm x)e_{01}(\bm x)}.
\]
For finite $\omega>0$, the four probabilities are obtained by solving for
\[
t(\bm x)=e_{11}(\bm x).
\]
Because
\[
e_{10}(\bm x)=p_0(\bm x)-t(\bm x),
\qquad
e_{01}(\bm x)=p_1(\bm x)-t(\bm x),
\]
and
\[
e_{00}(\bm x)=1-p_0(\bm x)-p_1(\bm x)+t(\bm x),
\]
the odds-ratio constraint becomes
\[
\omega
=
\frac{
t(\bm x)\{1-p_0(\bm x)-p_1(\bm x)+t(\bm x)\}
}{
\{p_0(\bm x)-t(\bm x)\}\{p_1(\bm x)-t(\bm x)\}
}.
\]
Thus, $t(\bm x)$ is selected as the feasible root in the interval
\[
\max\{0,p_0(\bm x)+p_1(\bm x)-1\}
\leq
t(\bm x)
\leq
\min\{p_0(\bm x),p_1(\bm x)\}.
\]
When both quadratic roots are numerically feasible, we choose the larger root,
which places $e_{11}(\bm x)$ closer to the upper boundary and minimizes
$e_{01}(\bm x)=p_1(\bm x)-e_{11}(\bm x)$. If neither root is feasible due to
numerical error, the solution is projected to the feasible interval.

When $\omega=1$, the construction reduces to conditional independence:
\[
e_{11}(\bm x)=p_0(\bm x)p_1(\bm x),
\]
\[
e_{10}(\bm x)=p_0(\bm x)\{1-p_1(\bm x)\},
\]
\[
e_{01}(\bm x)=\{1-p_0(\bm x)\}p_1(\bm x),
\]
and
\[
e_{00}(\bm x)=\{1-p_0(\bm x)\}\{1-p_1(\bm x)\}.
\]
When $\omega=\infty$, we used the monotonicity-limit construction by setting
$e_{10}(\bm x)=0$ whenever this is feasible. Thus, if
$p_1(\bm x)\geq p_0(\bm x)$,
\[
e_{11}(\bm x)=p_0(\bm x),
\qquad
e_{01}(\bm x)=p_1(\bm x)-p_0(\bm x),
\qquad
e_{00}(\bm x)=1-p_1(\bm x),
\qquad
e_{10}(\bm x)=0.
\]
If $p_1(\bm x)<p_0(\bm x)$, exact monotonicity is infeasible while preserving
both marginal probabilities. In this case, we projected to the boundary by
setting
\[
e_{01}(\bm x)=0,
\qquad
e_{11}(\bm x)=p_1(\bm x),
\qquad
e_{10}(\bm x)=p_0(\bm x)-p_1(\bm x),
\qquad
e_{00}(\bm x)=1-p_0(\bm x).
\]
For numerical stability, all four probabilities were truncated at zero if
necessary and renormalized to sum to one.

After computing
\[
\{e_{00}(\bm X_i),e_{01}(\bm X_i),e_{10}(\bm X_i),e_{11}(\bm X_i)\},
\]
we sampled the latent principal stratum
\[
(D_i(0),D_i(1))\mid \bm X_i
\sim
\mathrm{Multinomial}
\{1;
e_{00}(\bm X_i),
e_{01}(\bm X_i),
e_{10}(\bm X_i),
e_{11}(\bm X_i)
\}.
\]
The observed intermediate variable was then
\[
D_i=(1-Z_i)D_i(0)+Z_iD_i(1).
\]

We generated potential outcomes using nonlinear functions of the first five
covariates and the corresponding potential intermediate variable. Define
\[
\mathrm{xor}(u,v)
=
\mathbb I\{(u>0)\neq(v>0)\},
\qquad
\mathrm{and}(u,v)
=
\mathbb I(u>0,v>0).
\]
To limit the influence of extreme product terms, we used the capped variables
\[
x_j^c=\mathrm{cap}_{2.5}(x_j)
=
\min\{\max(x_j,-2.5),2.5\},
\qquad j=4,5.
\]
Define
\[
u_1=\sigma(1.2x_1),
\qquad
u_2=\sin(1.1x_2),
\qquad
u_3=\mathbb I(x_3>0),
\]
and the interaction terms
\[
I_{34}=x_3x_4^c,
\qquad
I_{35}=\mathrm{xor}(x_3,x_5),
\qquad
I_{45}=\mathrm{and}(x_4,x_5).
\]
The baseline outcome surface under control was
\begin{align*}
g_0(\bm x)
&=
-1
+1.4(u_1-\tfrac12)
+1.0u_2
+1.2(u_3-\tfrac12)
+0.6I_{34}
\\
&\quad
+0.9I_{35}(0.7+0.3|x_5^c|)
+0.7I_{45}(0.7+0.3|x_4^c|).
\end{align*}
The baseline outcome surface under treatment was
\begin{align*}
g_1(\bm x)
&=
-0.5
+1.2\{\sigma(1.1x_1)-\tfrac12\}
+1.1\sin(1.0x_2+0.3u_3)
+1.0\{\mathbb I(x_3>-0.2)-\tfrac12\}
\\
&\quad
+0.7x_3x_5^c
+0.8\,\mathrm{xor}(x_3,x_4)(0.7+0.3|x_4^c|)
+0.6\,\mathrm{and}(x_4,x_5)(0.7+0.3|x_5^c|).
\end{align*}
We further introduced $D$-dependent nonlinear modulation terms:
\[
s_0(\bm x)
=
0.8x_1x_5
-0.6\sin(x_2x_4)
+0.4(x_1-x_2)^2,
\]
and
\[
s_1(\bm x)
=
0.7x_4\mathbb I(x_3>0)
+0.5\cos(x_2x_5)
-0.4x_2^2.
\]
Thus,
\[
h_0(\bm x,d_0)=g_0(\bm x)+d_0s_0(\bm x),
\qquad
h_1(\bm x,d_1)=g_1(\bm x)+d_1s_1(\bm x).
\]
To satisfy conditional principal ignorability in the outcome model, we allowed
each potential outcome to depend only on the corresponding potential
intermediate variable. Specifically, $Y(0)$ depends on $D(0)$ but not on
$D(1)$, and $Y(1)$ depends on $D(1)$ but not on $D(0)$. With independent errors
\[
\varepsilon_{0i},\varepsilon_{1i}
\stackrel{\mathrm{i.i.d.}}{\sim}
N(0,1),
\]
we generated
\[
Y_i(0)
=
h_0(\bm X_i,D_i(0))
+
\varepsilon_{0i},
\]
and
\[
Y_i(1)
=
h_1(\bm X_i,D_i(1))
+
\varepsilon_{1i}.
\]
The observed outcome was
\[
Y_i=(1-Z_i)Y_i(0)+Z_iY_i(1).
\]

Under this construction,
\[
m_{11}(\bm x)
=
E(Y\mid Z=1,D=1,\bm X=\bm x)
=
E\{Y(1)\mid D(1)=1,\bm X=\bm x\}.
\]
Because $Y(1)$ depends on the principal stratum only through $D(1)$,
\[
m_{11}(\bm x)
=
g_1(\bm x)+s_1(\bm x).
\]
Similarly,
\[
m_{01}(\bm x)
=
E(Y\mid Z=0,D=1,\bm X=\bm x)
=
E\{Y(0)\mid D(0)=1,\bm X=\bm x\}
=
g_0(\bm x)+s_0(\bm x).
\]
Therefore, the true individual-level CSACE function is
\[
\mu_{11}(\bm x)
=
m_{11}(\bm x)-m_{01}(\bm x)
=
\{g_1(\bm x)+s_1(\bm x)\}
-
\{g_0(\bm x)+s_0(\bm x)\}.
\]

\subsection{Additional VIM Simulation Results}

\begin{figure}[htpb!]
\centering
\includegraphics[width=0.9\textwidth]{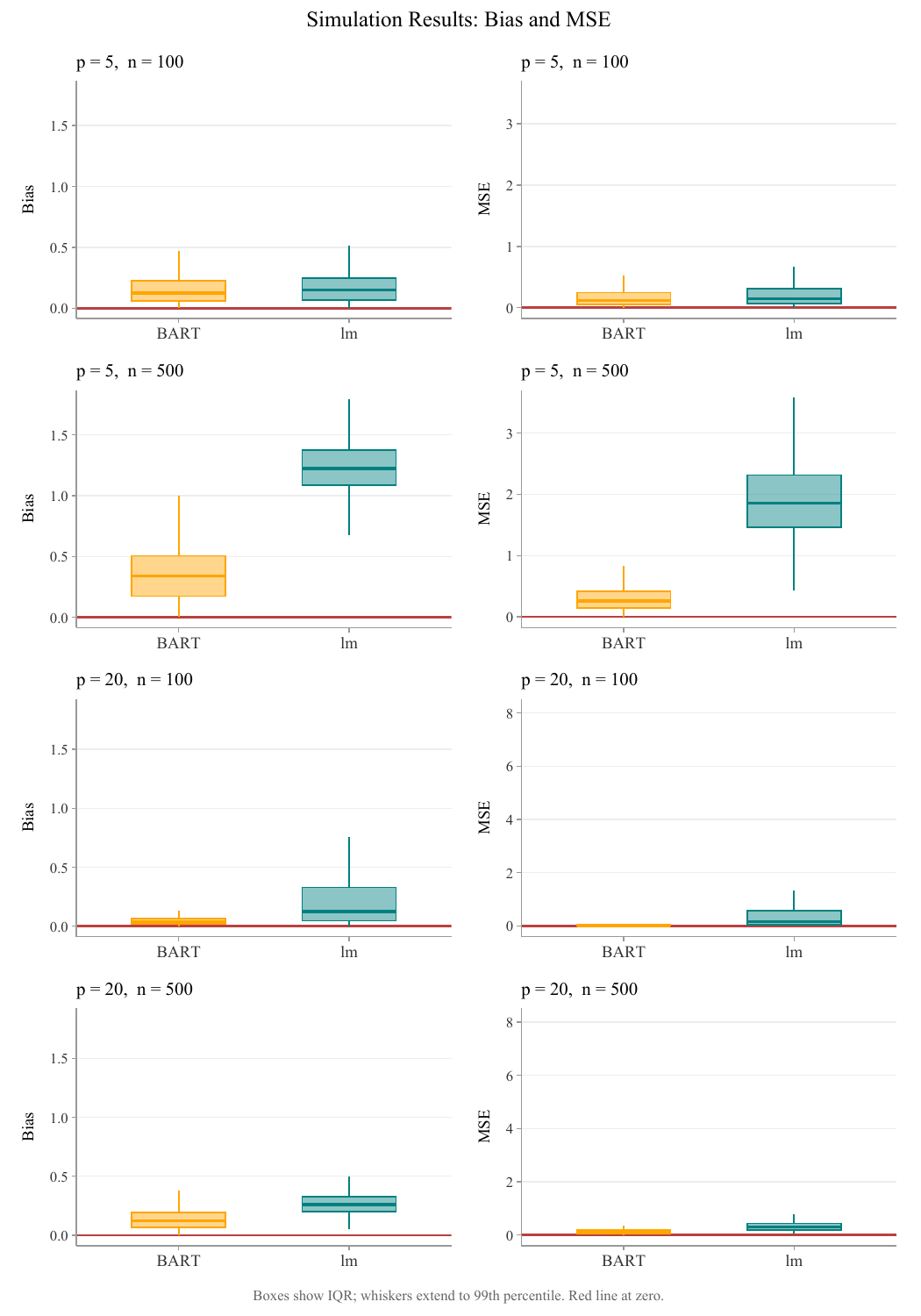}
\caption{Estimation accuracy of the CSACE-based variable importance measure in Experiment~2. Boxplots summarize the Monte Carlo distribution of VIM bias and MSE for BART and the parametric linear regression model across the four combinations of $p\in\{5,20\}$ and $n\in\{100,500\}$. The red horizontal line marks zero bias. BART yields smaller bias and MSE across all settings.}
\label{fig:supp_sim2_estimation}
\end{figure}

\begin{figure}[htpb!]
\centering
\includegraphics[width=\textwidth]{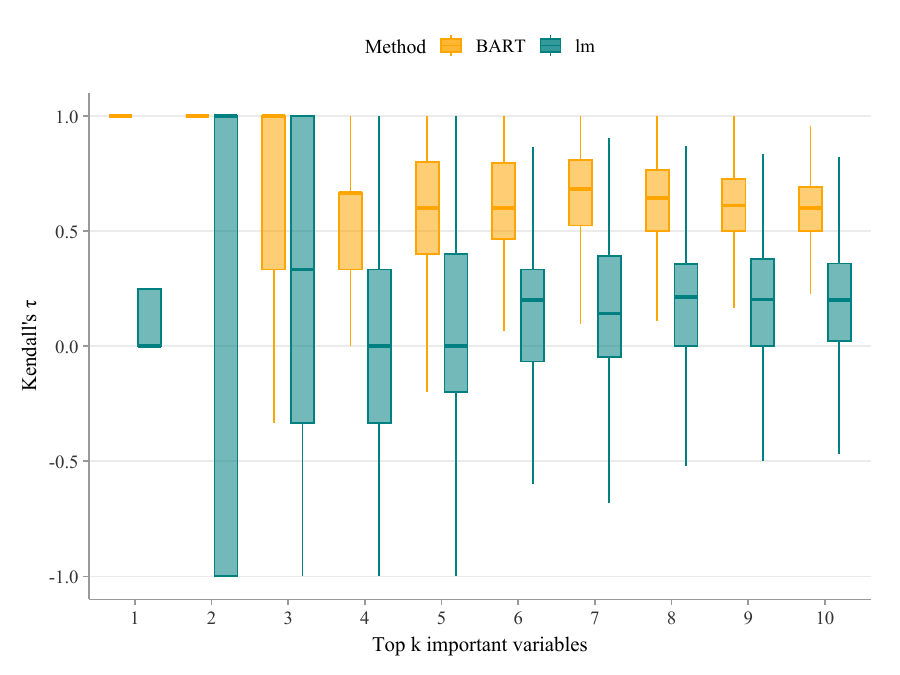}
\caption{Top-$k$ rank agreement between the estimated and oracle CSACE-based VIM rankings in Experiment~2. Agreement is summarized by Kendall's $\tau$ for $k\ge2$, with $k=1$ interpreted as exact recovery of the top-ranked variable. Larger values indicate better recovery of the oracle ordering. BART shows substantially stronger ranking performance than the parametric linear regression model, especially for the most important variables.}
\label{fig:supp_sim2_ranking}
\end{figure}

Supplementary Figures~\ref{fig:supp_sim2_estimation} and~\ref{fig:supp_sim2_ranking} provide the detailed results for Experiment~2 summarized in the main text.

\clearpage
\section{Technical Details for the Distribution-Aware Conditional Inference Tree}
\label{sec:supp-dacit}

For a node $t$ with index set $\mathcal I_t$ and $n_t=|\mathcal I_t|$, let $\bm\Delta_t=(\delta_{ij})_{i,j\in\mathcal I_t}$ denote the submatrix of pairwise distances between individualized posterior CSACE distributions. Its Gower-centered representation is
\[
\bm G_t=-\frac12\bm J_t\bm\Delta_t^{\circ2}\bm J_t,
\qquad
\bm J_t=\bm I_{n_t}-\frac1{n_t}\bm1_{n_t}\bm1_{n_t}^{\top}.
\]

Let $\mathcal V^{(t)}$ be the covariates available for splitting. For $X_j\in\mathcal V^{(t)}$, let $\bm H_t^{\mathrm{full}}$ be the projection matrix for the model containing all covariates in $\mathcal V^{(t)}$ and $\bm H_{t,-j}$ the projection matrix after excluding $X_j$. The marginal and residual sums of squares are
\[
SS_{t,j}=\operatorname{tr}\{(\bm H_t^{\mathrm{full}}-\bm H_{t,-j})\bm G_t\},
\qquad
SS_{t,E}=\operatorname{tr}\{(\bm I_{n_t}-\bm H_t^{\mathrm{full}})\bm G_t\}.
\]
The marginal PERMANOVA statistic is
\[
F_{t,j}=\frac{SS_{t,j}/q_{t,j}}{SS_{t,E}/(n_t-r_t)},
\]
where $q_{t,j}=\operatorname{rank}(\bm H_t^{\mathrm{full}})-\operatorname{rank}(\bm H_{t,-j})$ and $r_t=\operatorname{rank}(\bm H_t^{\mathrm{full}})$. Permuting the rows and columns of $\bm\Delta_t$ jointly yields $p_{t,j}$. This corresponds to marginal testing in the \texttt{adonis2} function of the R package \texttt{vegan}. If all $p_{t,j}>\alpha$, the node is terminal; otherwise $j_t^*=\arg\min_jp_{t,j}$.

For an admissible split $c$ of $X_{j_t^*}$, let $\bm z_{t,c}$ indicate membership in the left child. Continuous covariates use threshold splits $X_{ij_t^*}\leq c$, whereas categorical covariates use admissible subsets of levels. With $\bm Z_{t,c}=(\bm1_{n_t},\bm z_{t,c})$, the induced projection is
\[
\bm H_{t,c}=\bm Z_{t,c}(\bm Z_{t,c}^{\top}\bm Z_{t,c})^{-1}\bm Z_{t,c}^{\top}.
\]
Because $\bm G_t\bm1_{n_t}=\bm0$, the same partition can be represented by $\widetilde{\bm z}_{t,c}=\bm J_t\bm z_{t,c}$ and
\[
\widetilde{\bm H}_{t,c}=\widetilde{\bm z}_{t,c}
(\widetilde{\bm z}_{t,c}^{\top}\widetilde{\bm z}_{t,c})^{-1}
\widetilde{\bm z}_{t,c}^{\top}.
\]
The between- and within-child distributional variations are
\[
SS_{t,c}^{\mathrm{between}}=\operatorname{tr}(\widetilde{\bm H}_{t,c}\bm G_t),
\qquad
SS_{t,c}^{\mathrm{within}}=\operatorname{tr}\{(\bm I_{n_t}-\widetilde{\bm H}_{t,c})\bm G_t\}.
\]
Symmetry and idempotence of $\widetilde{\bm H}_{t,c}$ give the equivalent quadratic trace forms used in the main-text criterion:
\[
Q_t(c)=
\frac{\operatorname{tr}(\widetilde{\bm H}_{t,c}\bm G_t\widetilde{\bm H}_{t,c})}
{\operatorname{tr}\{(\bm I_{n_t}-\widetilde{\bm H}_{t,c})\bm G_t(\bm I_{n_t}-\widetilde{\bm H}_{t,c})\}}.
\]
All candidate splits at a node have one between-child degree of freedom and the same residual degrees of freedom, so maximizing $Q_t(c)$ is equivalent to maximizing the corresponding PERMANOVA pseudo-$F$ statistic. After selecting $c_t^*=\arg\max_cQ_t(c)$, DaCIT removes $X_{j_t^*}$ from both child-node candidate sets and continues recursively subject to the stopping rules.

\section{Additional Results of Real Data Analysis}
This section provides additional results from the real-data analysis of the ARDS Network higher- versus lower-PEEP trial. In the main text, we focused on the leading variables identified by the CSACE variable-importance measure and on the primary distribution-aware subgroup tree used to summarize heterogeneity in the conditional survivor average causal effect. Here, we present two supplementary analyses. First, we report the complete list of CSACE-VIM estimates across all baseline covariates, which provides a fuller view of how different demographic, physiologic, and ventilatory characteristics contributed to heterogeneity in the estimated survivor causal effect. Second, we assess the robustness of the subgroup-discovery results through a sensitivity analysis of the tree-construction procedure. The results are summarized below.

\begin{table}[b!]
\centering
\caption{Complete CSACE-VIM results for the ARDS trial}
\label{tab:supp_csace_vim}
\centering
\begin{tabular}[t]{llr}
\toprule
Variable & Description & $\xi = 0.25$\\
\midrule
AGE & Age at enrollment & 8.374\\
GENDER & Female or Male & 3.523\\
HEIGHTC & Height used to derive predicted body weight for tidal-volume target & 1.792\\
PSTATVC & Plateau pressure recorded for the volume-control (VC) context & 1.466\\
ARTPH & Arterial blood pH from arterial blood gas & 0.931\\
\addlinespace
FIO2 & Fraction of inspired oxygen at the time of measurement & 0.662\\
TRESPR & Total respiratory rate & 0.435\\
HRATE & Heart rate & 0.363\\
MAPRES & Mean airway pressure recorded during mechanical ventilation & 0.346\\
TMNVNT & Total minute ventilation & 0.335\\
\addlinespace
SPO2 & Oxyhemoglobin saturation measured by pulse oximetry & 0.303\\
PACO2 & Partial pressure of arterial carbon dioxide & 0.298\\
PEAK & Peak inspiratory pressure & 0.295\\
SYSBP & Systolic blood pressure & 0.292\\
TIDALVC & Calculated delivered tidal volume in the VC context & 0.291\\
PSTAT1 & Plateau pressure measured as end-inspiratory plateau & 0.170\\
\addlinespace
DIABP & Diastolic blood pressure & 0.164\\
PEEP & Positive end-expiratory pressure & 0.132\\
PAO2 & Partial pressure of arterial oxygen & 0.113\\
TIDAL & Calculated delivered tidal volume & 0.072\\
VASOL242 & Use of vasopressors within the last 24 hours prior to enrollment & 0.064\\
\addlinespace
ASPIR0 & Indicator for aspiration as the cause of lung injury, 0 is No & 0.048\\
PCON1 & Ventilator mode indicator: Pressure Control & 0.045\\
ASPIR2 & Indicator for aspiration as the cause of lung injury, 2 is Unknown & 0.043\\
SRATE & Set ventilator respiratory rate & 0.022\\
ASPIR1 & Indicator for aspiration as the cause of lung injury, 1 is Yes & 0.005\\
\bottomrule
\end{tabular}
\end{table}

Table~\ref{tab:supp_csace_vim} reports the complete CSACE-VIM results for all baseline covariates included in the real-data analysis. Consistent with the main-text findings, age had the largest variable-importance value, indicating that it was the dominant contributor to heterogeneity in the estimated conditional survivor average causal effect. Sex was the second most important variable, followed by height, plateau pressure in the volume-control context, and arterial pH. Several physiologic and ventilatory variables, including fraction of inspired oxygen, total respiratory rate, heart rate, mean airway pressure, minute ventilation, oxygen saturation, arterial carbon dioxide pressure, peak inspiratory pressure, and systolic blood pressure, also showed non-negligible contributions. In contrast, variables related to aspiration status, ventilator mode, set ventilator respiratory rate, and positive end-expiratory pressure had relatively small CSACE-VIM values. Overall, the complete ranking suggests that heterogeneity in the survivor causal effect was concentrated in demographic characteristics, body-size-related ventilatory targeting, and markers of respiratory or physiologic status.

\begin{table}[htpb!]
\centering
\caption{Identified subgroups under different $\xi$ values. $n$ denotes subgroup size; $p$ denotes the posterior probability that higher PEEP is beneficial.}
\label{tab:subgroups}
\resizebox{\textwidth}{!}{%
\begin{tabular}[t]{cll}
\toprule
$\xi$ & \textbf{Beneficial Subgroup} & \textbf{Harmful Subgroup} \\
\midrule
$-0.5$ & Age$\leq$55, Female, ARTPH$>$7.355 & Male, Age$>$59.5, Height$\leq$166.3 \\
        & $n=53$, $p=0.995$ & $n=11$, $p=0.019$ \\
\addlinespace
$-0.25$ & Age$\leq$55, Female, ARTPH$>$7.355 & Male, Age$>$65.5 \\
         & $n=54$, $p=0.995$ & $n=44$, $p=0.021$ \\
\addlinespace
$0$ & Age$\leq$55, Female, ARTPH$>$7.355 & Male, Age$>$59.5, Height$\leq$166.3 \\
    & $n=55$, $p=0.995$ & $n=11$, $p=0.019$ \\
\addlinespace
$0.25$ & Age$\leq$54.5, Female, ARTPH$>$7.335 & Male, Age$>$65.5 \\
        & $n=62$, $p=0.993$ & $n=45$, $p=0.023$ \\
\addlinespace
$0.5$ & Age$\leq$54.5, Female, ARTPH$>$7.335 & Age$>$54.5, ARTPH$\leq$7.325, PSTATVC$>$31.5 \\
      & $n=64$, $p=0.993$ & ($n=10$, $p=0.003$); PSTATVC$\leq$31.5 ($n=13$, $p=0.029$) \\
\addlinespace
$0.75$ & Age$\leq$54.5, Female, ARTPH$>$7.335 & Age$>$54.5, ARTPH$\leq$7.325, PSTATVC$>$31.5 \\
        & $n=64$, $p=0.993$ & ($n=10$, $p=0.003$); PSTATVC$\leq$31.5 ($n=13$, $p=0.029$) \\
\addlinespace
$1$ & Age$\leq$54.5, Female, ARTPH$>$7.335 & Age$>$54.5, ARTPH$\leq$7.325, PSTATVC$>$31.5 \\
    & $n=65$, $p=0.993$ & ($n=10$, $p=0.003$); PSTATVC$\leq$31.5 ($n=13$, $p=0.029$) \\
\addlinespace
$1.25$ & Age$\leq$54.5, Female, ARTPH$>$7.335 & Age$>$54.5, ARTPH$\leq$7.325, PSTATVC$>$31.5 \\
        & $n=65$, $p=0.993$ & ($n=10$, $p=0.003$); PSTATVC$\leq$31.5 ($n=13$, $p=0.029$) \\
\addlinespace
$1.5$ & Age$\leq$55.5, Female, ARTPH$>$7.335 & Male, Age$>$59.5, Height$\leq$166.3 \\
      & $n=66$, $p=0.995$ & ($n=11$, $p=0.019$); Female, Age$>$55.5, ARTPH$\leq$7.335 ($n=17$, $p=0.026$) \\
\bottomrule
\end{tabular}
}
\end{table}

Table~\ref{tab:subgroups} summarizes the sensitivity analysis for subgroup discovery under different values of \(\xi\), where \(\xi\) controls the log-odds association between the two potential survival statuses. A positive \(\xi\) corresponds to stronger concordance between $D(0)$ and $D(1)$, and therefore implies that individuals who would survive under one treatment are more likely to also survive under the other treatment. Across the full range of \(\xi\) values, the beneficial subgroup was highly stable: it was consistently characterized by younger female patients with relatively preserved arterial pH. The recurring sex-related split is clinically hypothesis-generating. Sex and height determine predicted body weight and thus influence tidal-volume targets, and recent work has questioned whether commonly used formulas generalize equally across patient groups and whether lung-protective ventilation is sex neutral~\citep{sarma2025evaluating,chiumello2026lung}. These considerations may contribute to heterogeneity in delivered ventilation, but they do not establish that sex modifies the effect of higher PEEP in this trial. Younger age may reflect greater physiologic reserve and potential to benefit from a higher-PEEP strategy. The age cutoff varied only slightly, and the subgroup size and posterior probability of benefit remained nearly unchanged. This stability suggests that the evidence for benefit in this subgroup is driven mainly by the posterior outcome contrast among estimated always-survivors, rather than by small changes in the assumed latent principal-stratum membership model. Clinically, this is also consistent with the main analysis, in which younger female patients with better acid--base status formed the clearest subgroup with shorter time to discharge under higher PEEP.

In contrast, the harmful subgroup showed greater sensitivity to \(\xi\). Some features were nevertheless recurrent: older age appeared repeatedly across nearly all values of \(\xi\), and male patients were often represented in the harmful subgroup when \(\xi\) was small or moderate. A more noticeable structural change occurred beginning around \(\xi=0.5\), where the harmful subgroup shifted from being primarily defined by male sex and older age to being defined by older age combined with low arterial pH and plateau pressure. Low pH may identify shock or severe respiratory or metabolic compromise, whereas high plateau pressure may indicate limited capacity to increase end-inspiratory lung volume without over-distention. In such patients, higher PEEP could impose greater hemodynamic or over-distention-related costs~\citep{sahetya2017fifty,luecke2005clinical,xingzheng2024impact}. This pattern is interpretable because increasing \(\xi\) strengthens the assumed concordance between survival under lower and higher PEEP, which changes the posterior allocation of patients to the always-survivor stratum. As the estimated always-survivor population becomes more concentrated among patients with similar potential survival statuses, harm signals may become more strongly tied to physiologic severity markers rather than to demographic splits alone. These mechanism-based explanations remain hypotheses rather than findings directly tested by the subgroup analysis. Thus, the sensitivity analysis supports a stable exploratory beneficial subgroup while indicating that the precise characterization of harmful subgroups is more dependent on the sensitivity parameter and should be interpreted more cautiously.

\bibliographystyle{imsart-nameyear}
\bibliography{bibliography.bib}